\documentclass[journal,comsoc]{IEEEtran}

\usepackage{graphicx}
\ifCLASSINFOpdf
\else
\fi

\usepackage{amsmath}
\usepackage{verbatim}
\usepackage{multirow}

\usepackage{caption}
\usepackage{subcaption}

\usepackage[cmintegrals]{newtxmath}
\usepackage{algorithmic}
\usepackage{algorithm}

\usepackage{mathtools}

\usepackage{array}

\begin{document}

\title{Marine IoT Systems with Space-Air-Sea Integrated Networks: Hybrid LEO and UAV Edge Computing}

\author{Sooyeob~Jung,~Seongah~Jeong,~Jinkyu~Kang,~and~Joonhyuk~Kang 
\vspace{-25pt}
\thanks{This work was supported by the Institute of Information \& communications Technology Planning \& Evaluation (IITP) grant funded by the Korea government (MSIT) (No.2021-0-00847, Development of 3D Spatial Satellite Communications Technology).}
\thanks{This research was supported by the Ministry of Science and ICT (MSIT), Korea, under the Information Technology Research Center (ITRC) support program (IITP-2020-0-01787) supervised by the IITP.}
\thanks{Sooyeob Jung is with the Department of Electrical Engineering, Korea Advanced Institute of Science and Technology (KAIST), and with the Satellite Wide-Area Infra Research Section, Electronics and Telecommunications Research Institute (ETRI), Daejeon, South Korea (Email: jung2816@kaist.ac.kr).}
\thanks{Seongah Jeong is with the School of Electronics Engineering, Kyungpook National University, Daegu 14566, South Korea (Email: seongah@knu.ac.kr).}
\thanks{Jinkyu Kang is with the Department of Information and Communications Engineering, Myongji University, Gyeonggi-do 17058, South Korea (Email: jkkang@mju.ac.kr).}
\thanks{Joonhyuk Kang is with the Department of Electrical Engineering, Korea Advanced Institute of Science and Technology (KAIST), Daejeon, South Korea (Email: jhkang@ee.kaist.ac.kr).}}

\maketitle

\begin{abstract} Marine Internet of Things (IoT) systems have grown substantially with the development of non-terrestrial networks (NTN) via aerial and space vehicles in the upcoming sixth-generation (6G), thereby assisting environment protection, military reconnaissance, and sea transportation. Due to unpredictable climate changes and the extreme channel conditions of maritime networks, however, it is challenging to efficiently and reliably collect and compute a huge amount of maritime data. In this paper, we propose a hybrid low-Earth orbit (LEO) and unmanned aerial vehicle (UAV) edge computing method in space-air-sea integrated networks for marine IoT systems. Specifically, two types of edge servers mounted on UAVs and LEO satellites are endowed with computational capabilities for the real-time utilization of a sizable data collected from ocean IoT sensors. Our system aims at minimizing the total energy consumption of the battery-constrained UAV by jointly optimizing the bit allocation of communication and computation along with the UAV path planning under latency, energy budget and operational constraints. For availability and practicality, the proposed methods were developed for three different cases according to the accessibility of the LEO satellite, ``Always On," ``Always Off" and ``Intermediate Disconnected", by leveraging successive convex approximation (SCA) strategies. Via numerical results, we verify that significant energy savings can be accrued for all cases of LEO accessibility by means of joint optimization of bit allocation and UAV path planning compared to partial optimization schemes that design for only the bit allocation or trajectory of the UAV.

\\
\end{abstract}

\IEEEpeerreviewmaketitle

\section{Introduction}
\IEEEPARstart{M}{arine} Internet of Things (IoT) systems have evolved significantly with the rapid development of non-terrestrial network (NTN) technologies composed of space and airborne platforms to collect and process a variety of ocean data. The vast amount of ocean data plays an important role in marine monitoring, which contributes to environmental protection, natural disaster prevention, oceanographic research, mineral exploration, military surveillance, etc. [\ref{ocean3}]-[\ref{ocean2}]. In particular, continuous monitoring of various physical phenomena of marine networks, such as sounds, vibrations and images, requires high-precision and wide-range measurements. Currently, three types of marine monitoring platforms are being investigated according to the relay node: shore-based radar, survey vessels and satellites [\ref{ocean3}], most of which have the following procedures. By using existing information communication technologies, the marine data collected from ocean IoT sensors is transferred to a ground cloud server with sufficient computation storage capacity. The ground cloud server stores and analyzes the collected data, thereby managing various applications based on ocean utilization and exploration. In shore-based radar systems installed on offshore buoys and automatic weather stations located on the coast or islands, there are difficulties in installation and maintenance due to their spatial constraints. Meanwhile, survey vessel-based platforms have temporal constraints, which limit the time for data collection. In addition, unexpected loss and defects of collected data may occur in point measurements attained by platforms with shore-based radar or survey vessel platforms due to extreme channel environments and unpredictable climate changes in the ocean [\ref{ocean}].

To address these spatial and temporal limitations, satellite-based monitoring can be an alternative that provides full coverage of the area of interest with one or multiple satellites. With the participation of global companies in the satellite business such as SpaceX, Amazon, and Telesat [\ref{LEO}], low-Earth orbit (LEO) satellites are gaining more attention than ever before, and cost-effective easy-to-deploy large-scale satellite networks are being established. In addition, conventional satellite operators such as Spire, Kepler, Fleet, Lacuna space and Eutelsat, are preparing to provide satellite IoT services with global coverage [\ref{satIoT}], [\ref{satIoT1}]. Until recently, satellites have mostly been adopted as a relay with terrestrial networks; however, for future 6G IoT services, they can operate as functional network components, e.g., computing servers [\ref{Sat-Com2}]-[\ref{SatEC1}]. Traditionally, the critical drawback of satellite-assisted networks is the latency resulting from round-trip delays due to the IoT sensor-satellite-terrestrial station link as well as the rapidly increasing volume of transmitted data. Therefore, it is beneficial to bring computing functions in the satellite to handle processing capabilities of the collected data, rather than sending it to the ground cloud server. In the following section, we briefly summarize the recent research activities that focus on hierarchical integrated networks using satellites as computing servers.

\subsection{Related Works}

Satellite-assisted edge computing systems have been actively studied in space-ground integrated networks [\ref{Sat-Com2}]-[\ref{Sat-Com1}], space-air-ground integrated networks (SAGIN) [\ref{Nan}]-[\ref{SAGIN}] and space-air-sea-based non-terrestrial networks (SAS-NTN) [\ref{SASIN1}], [\ref{SASIN2}]. In particular, the authors in [\ref{Sat-Com2}] propose a three-tier computation architecture consisting of ground users, LEO satellites and ground servers to minimize the total energy consumption of the system. In [\ref{Sat-Com3}], network slice scheduling for satellite-assisted computing architecture is studied, where satellite servers and ground servers are considered for IoT applications. Although satellite-assisted edge computing can provide real-time offloading services to large areas, such as the ocean, it still faces several practical problems. For long-distance communication with a satellite, more transmit power and larger antenna size are preferred at ground user terminals, which is costly and spatially-limited in real applications. Moreover, the transceiver for satellite communications must be robustly designed against severe fading due to atmospheric turbulence.

Unmanned aerial vehicles (UAVs) can be adopted to provide enhanced coverage for overcoming path loss and fading issues of satellite-assisted edge computing. UAVs can receive and compute data in close proximity to ocean IoT sensors, or can relay the data to the cloud server for computing. Recently, UAV-assisted satellite IoT networks have been suggested in several studies [\ref{Nan}], [\ref{channel2}]. Cheng \emph{et al.} [\ref{Nan}] propose offloading systems of remote IoT applications in the space-air-ground scenario, where UAVs provide computational capability to nearby users as edge servers, while satellites relay the offloaded data to the ground cloud server. In [\ref{channel2}], LEO satellite-assisted UAV data collection for IoT sensors is proposed, where the delay-tolerant data and delay-sensitive data are transferred to the ground cloud server via UAV and LEO satellite, respectively.

As briefly reviewed above, most of existing works on hierarchical offloading systems in the integrated space and air networks assume terrestrial infrastructures, which may result in latency caused by the extreme channel variation of marine IoT systems. Furthermore, even though space or aerial computing platforms are considered, most studies assume full accessibility of the LEO satellite during mission time, which may not be guaranteed according to the orbit of revolution of the LEO satellite under insufficient deployments. To perform real-time data mining and analysis of ocean data in marine IoT systems, the use of aerial/space moving cloudlets play an important role considering their availability.

\begin{figure}[t]\label{system}
    \includegraphics[width=\columnwidth]{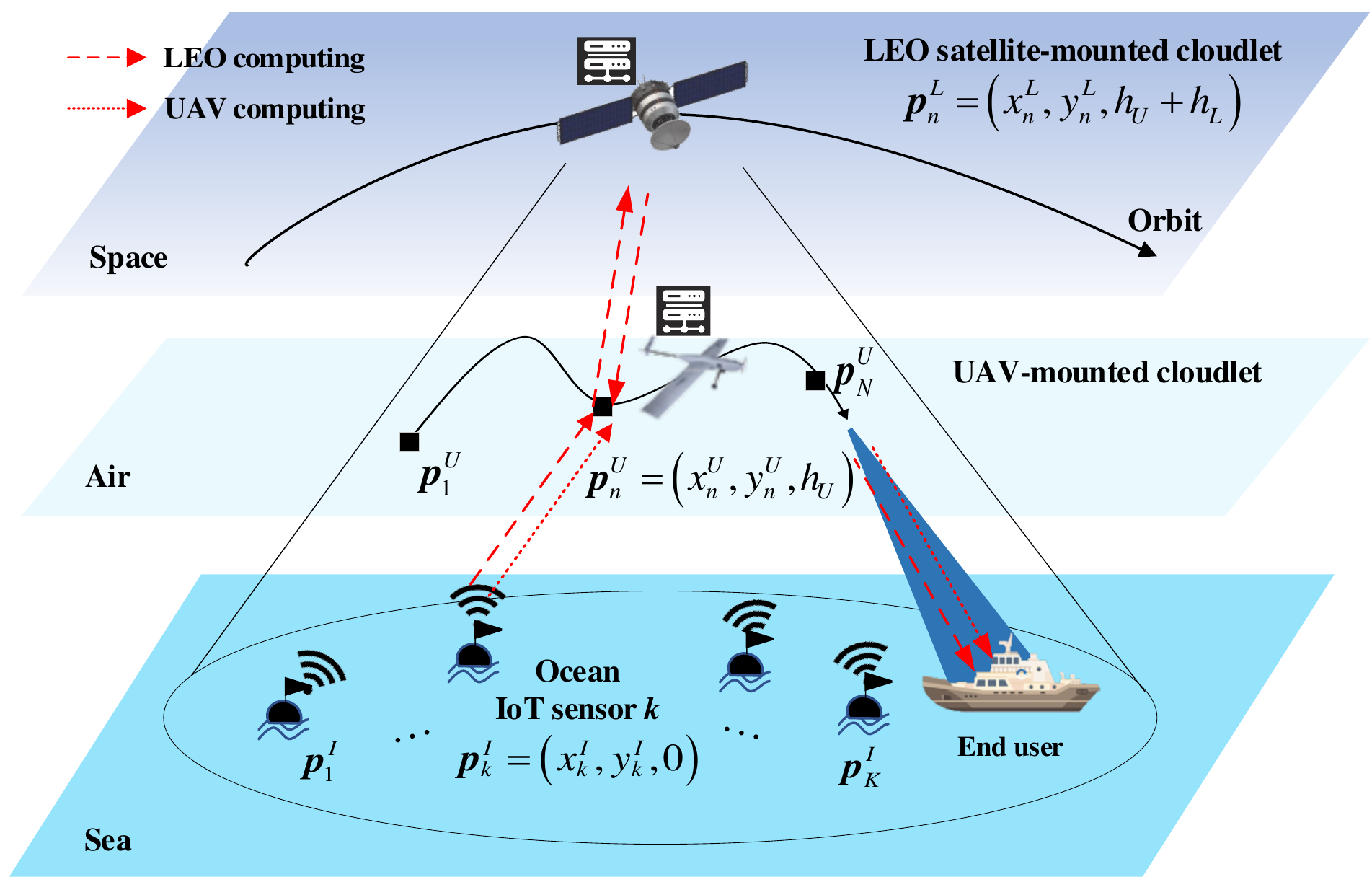}
    \caption{Marine IoT system model with a space-air-sea integrated network using hybrid LEO and UAV edge computing for real-time data utilization.}
\end{figure}

\subsection{Main Contributions}

In this paper, we focus on a marine IoT system with space-air-sea integrated networks, as illustrated in Fig. 1, where both UAV and LEO satellite-mounted cloudlets are deployed to offer computing opportunities. In the proposed system, a number of ocean IoT sensors are distributed only to collect abundant marine information with limited battery, and transmit the collected data to a designated computing server among UAV or LEO-mounted cloudlets so as to satisfy the system design criterion. Here, the LEO satellite is assumed to have a higher computational capability to process the task than that of the UAV. When the IoT data size exceeds the computation capacity of the UAV, the computational task is totally offloaded to the LEO satellite. The computation results executed at LEO are retransmitted to the UAV, are stored until it arrives over the end user, and is finally sent to the end user. To this end, we tackle the key design problem of jointly optimizing the bit allocation for communication and computing and the trajectory of the UAV, with the aim of minimizing its energy consumption. The main contributions of this paper are summarized as follows:

\begin{itemize}
\item For marine IoT systems with extreme channel environments and unpredictable climate changes, we propose a hybrid LEO and UAV edge computing method. The scheduling between UAV and LEO satellite-mounted cloudlets depends on the size of the offloaded ocean data and the LEO connection status.

\item For practicality and usability, we consider three different scenarios according to LEO availability such as ``Always On," ``Always Off" and ``Intermediate Disconnected". For each case, we develop the joint optimization of bit allocation required for offloading and UAV path planning.

\item The non-convex optimization problems formulated for three different cases depending on the availability of the LEO satellite are tackled by means of a successive convex approximation (SCA) algorithm [\ref{SCA1}], [\ref{SCA2}], which can guarantee the local minimum of the original non-convex problems by using an efficient iterative algorithm. 
\end{itemize}

The rest of this paper is organized as follows. The system model is presented in Section II. Section III, IV and V provide problem formulations and proposed methods for the LEO access status of ``Always On," ``Always Off" and ``Intermediate Disconnected", respectively. Simulation results are given in Section VI, and conclusions are summarized in Section VII.

\section{System Model}

\subsection{Set-up}

Fig. 1 illustrates a marine IoT system with a space-air-sea integrated network using hybrid LEO and UAV edge computing, where $K$ ocean IoT sensors collect marine data to be entirely transferred to available cloudlets for computing. The computed results are then designated to an end user. For real-time data utilization, two types of cloudlets mounted on the UAV and LEO satellite are considered, between which the scheduling depends on the UAV computing capability and LEO accessibility. Specifically, when the collected data size exceeds the computation capacity of the UAV, the data should be entirely offloaded to the LEO. The computing capability of the LEO satellite is assumed to be higher than that of the UAV. Another major factor for scheduling is whether the LEO satellite is available or not since its beam coverage varies according to the orbit of revolution. Here, we consider three different cases according to the availability of the LEO satellite during mission time: ``Always On," ``Always Off" and ``Intermediate Disconnected". For each scenario, we developed the joint optimization of the bit allocation for communication and computation and the trajectory of the UAV. Depending on the types of cloudlets, we refer to \textit{UAV computing} and \textit{LEO computing}, where computing of the IoT sensor task is executed at the UAV and LEO, respectively. In UAV computing, the task of the IoT sensor $k$ is offloaded to the UAV-mounted cloudlet until the UAV arrives over the end user and the output results are conveyed to them. In LEO computing, the UAV receives and relays the offloaded data of the IoT sensor to LEO for the LEO execution. The computed results at LEO are then sent to the end user via the UAV when the UAV arrives above them.

 \begin{figure}[t]
    \includegraphics[width=\columnwidth]{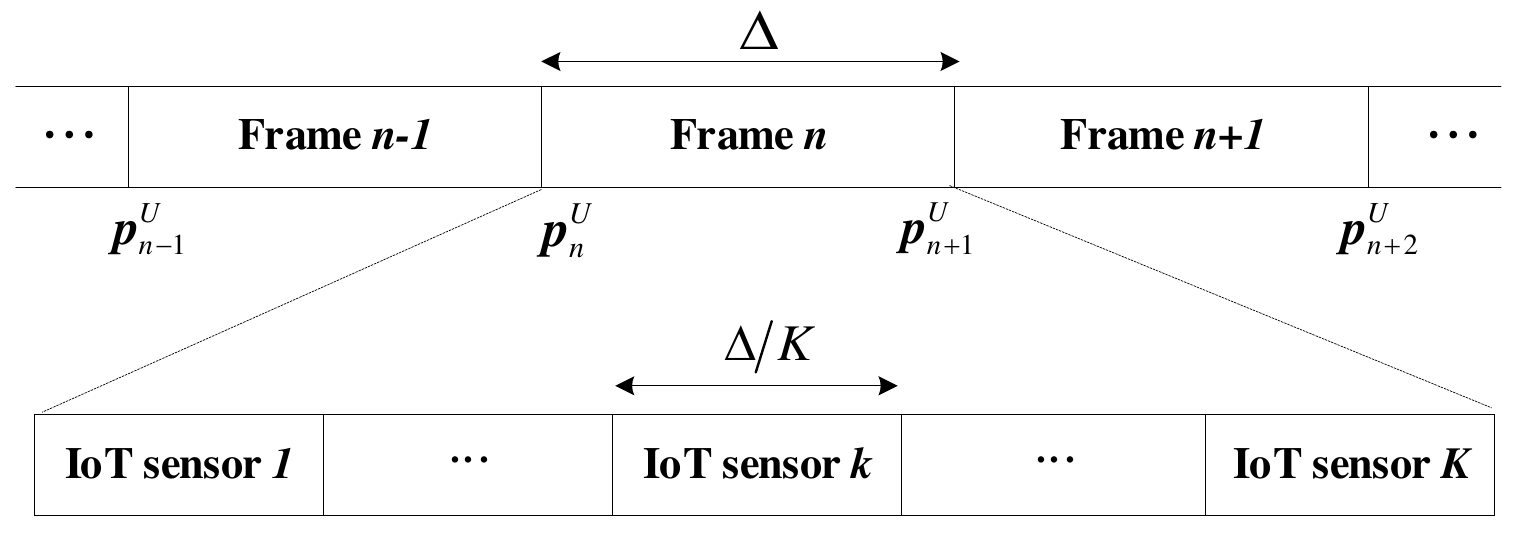}
    \caption{Frame structure of orthogonal access for multiple ocean IoT sensors.}
\end{figure}

\begin{table}[t]
    \caption{List of Symbols}\label{Table}
    \centering
    \begin{tabular}{|c|| p{6.3cm} |}
    \hline
    Symbol & Definition  \\
    \hline 
    $K$         & Number of ocean IoT sensors \\
    \hline
    $T$         & Total mission time\\
    \hline
    $\Delta$    & Frame duration \\  
    \hline
    $N$         & Number of frames within $T$ \\  
    \hline
    $h_U, h_L$  & Altitudes of UAV and LEO satellite with respect to average sea surface level and UAV, respectively \\
    \hline
    ${g_{k,n}}, {h_{n}}$ & Path loss between the IoT sensor $k$ and UAV and between the UAV and LEO at the $n$th frame \\
    \hline
    $g_0$       & Channel gain at reference distance 1 m \\
    \hline
    $T_v$       & Visible time of an LEO satellite\\
    \hline
    $v_s$       & Speed of an LEO satellite \\
    \hline
    $h$         & Height of an LEO satellite orbit \\
    \hline
    $\theta, \varphi$   & Elevation angle and beamwidth of the LEO satellite\\
    \hline
    $M$         & the gross mass of the UAV \\
    \hline
    \rule{0pt}{8pt} ${\boldsymbol{v}_n^U}$   & velocity vector of the UAV at the $n$th frame \\ 
    \hline
    $\varepsilon$   & Energy budget of the IoT sensor $k$ at each frame \\
    \hline
    $I_k$     & Number of input bits of the IoT sensor $k$ \\   
    \hline
    \rule{0pt}{8.5pt} $E_{k,n}^{I,U}$   & Energy consumption for uplink communication at the IoT sensor $k$ at the $n$th frame\\
    \hline 
    \rule{0pt}{8.5pt} $E_{k,n}^U$, $E_{k,n}^{U,L}$ & Energy consumption for computing and uplink communication at the UAV-mounted cloudlet for the IoT sensor $k$ at the $n$th frame \\
    \hline
    \rule{0pt}{8pt} $E^{U,E}$ & Energy consumption for downlink communication at the UAV-mounted cloudlet\\ 
    \hline
    \rule{0pt}{8.5pt} $E_{k,n}^L$, $E_{k,n}^{L,U}$ & Energy consumption for computing and downlink communication at the LEO-mounted cloudlet for the IoT sensor $k$ at the $n$th frame \\
    \hline
    \rule{0pt}{8pt} $E_{n}^F$   & Energy consumption for a UAV flying at the $n$th frame \\
    \hline
    \rule{0pt}{8.5pt} ${L_{k,n}^{I,U}}$ & Number of bits for uplink communication at the IoT sensor $k$ at the $n$th frame\\
    \hline
    \rule{0pt}{8.5pt} ${l_{k,n}^{U}}, {L_{k,n}^{U,L}}$ & Number of bits for computing and uplink communication at a UAV-mounted cloudlet for the IoT sensor $k$ at the $n$th frame\\    
    \hline
    \rule{0pt}{8pt} ${L^{U,E}}$ & Number of bits for downlink communication at the UAV-mounted cloudlet\\    
    \hline
    \rule{0pt}{8.5pt} ${l_{k,n}^{L}}, {L_{k,n}^{L,U}}$ & Number of bits for computing and downlink communication at the LEO-mounted cloudlet for the IoT sensor $k$ at the $n$th frame\\    
    \hline
    \rule{0pt}{8pt} $O_k^L, O_k^U$     & Number of output bits produced per input bit of the IoT sensor $k$ \\
    \hline
    \rule{0pt}{8pt} ${f_n^L}, {f_n^U}$ & CPU frequency at the LEO and UAV-mounted cloudlets for the $n$th frame \\
    \hline
    \rule{0pt}{8pt} ${C_k^L}, {C_k^U}$ & CPU cycles per input bit at the LEO and UAV-mounted cloudlets for the task of the IoT sensor $k$ \\
    \hline
    \rule{0pt}{8pt} ${{\gamma ^L}}, {{\gamma ^U}}$ & Effective switched capacitances of the LEO and UAV, respectively\\ 
    \hline
    \rule{0pt}{8pt} $\boldsymbol{p}_k^I, \boldsymbol{p}_n^U, \boldsymbol{p}_n^L$ & Positions of the IoT sensor $k$, UAV and LEO for the $n$th frame \\
    \hline
    ${\alpha _{k,n}, \beta _{k,n}}$   & Variables to indicate LEO connection and offloading scheduling of the IoT sensor $k$ at the $n$th frame\\
    \hline
    $N_t$   & Frame number during LEO disconnection\\
    \hline
    \end{tabular}
\end{table}

For communication links between IoT sensors and the UAV, and between the UAV and LEO satellite, a frequency division duplex (FDD) scheme is assumed with equal bandwidth $B$ for the uplink and downlink. Each IoT sensor $k$ has the number $I_k$ of input information bits to be processed. The results for LEO computing and UAV computing are characterized as the number $O_k^L$ and $O_k^U$ of bits produced per input bit of the IoT sensor $k$, and the number ${C_k^L}$ and ${C_k^U}$ of CPU cycles per input bit for computing, respectively. We assume that all tasks must be computed within the total mission time $T$. Here, a three-dimensional Cartesian coordinate system is adopted based on the metric unit. We assume that the IoT sensor $k$ is deployed at position $\boldsymbol{p}_k^I = ( {x_k^I,y_k^I,a_k} )$, for $k \in \left\{{1, \cdot  \cdot  \cdot, K+1} \right\}$, with $a_k$ being the average sea surface level, where the position of the end user is considered with an index of $K+1$. The UAV flies along a trajectory ${\boldsymbol{p}^U}(t) = ( {{x^U}(t),{y^U}(t),{h_U}})$ with a fixed altitude $h_U$ assumed for system stability, for $0 \le t \le T$, and the position of the LEO satellite is defined as ${\boldsymbol{p}^L}(t) = ({{x^L}(t),{y^L}(t),{h_U+h_L}})$ with a fixed altitude $h_U + h_L$, for $0 \le t \le T$, all the altitudes are measured with respect to the average sea surface level $a_k$. For the multiple access of $K$ ocean IoT sensors, orthogonal access is assumed, as shown in Fig. 2. For tractability, in this paper, the total time duration $T$ is divided into $N$ frames of duration $\Delta$ seconds, each of which is equally divided as $\Delta /{K}$ seconds, and is preallocated to IoT sensors for uplink and downlink communication required for offloading. Accordingly, the IoT sensors do not interfere with each other in the offloading procedure. Moreover, the information data collected from the IoT sensor $k$ at the $n$ th frame is assumed to be entirely computed and transferred to the designated node within the corresponding frame during $\Delta /{K}$ seconds, for $n \in \left\{{1, \cdot  \cdot  \cdot, N}\right\}$, so that the computational task cannot be partitioned. According to the discretized time unit, the trajectory of the UAV ${\boldsymbol{p}^U}(t)$ and the position of the LEO satellite ${\boldsymbol{p}^L}(t)$ is expressed as $\boldsymbol{p}_n^U = ( {x_n^U,y_n^U,h_U})$ and $\boldsymbol{p}_n^L = ( {x_n^L,y_n^L,h_U+h_L} )$, for $n \in \mathcal{N}$, respectively. The LEO satellite generally flies at a constant speed along its orbit and the relative positional coordinates of the LEO and UAV should vary constantly. For the task mission of marine IoT systems, the initial location $\boldsymbol{p}_I^U$ and the final location $\boldsymbol{p}_F^U$ of the UAV are assigned to $\boldsymbol{p}_1^U$ and $\boldsymbol{p}_{N+1}^U$, respectively, and its maximum speed constraint is given as 
 \begin{align}
  \left\| {\boldsymbol{v}_n^U} \right\| = \frac{{\left\| {\boldsymbol{p}_{n + 1}^U - \boldsymbol{p}_n^U} \right\|}}{{\Delta }} \le {v_{\max}}, 
 \end{align}
where the velocity vector ${\boldsymbol{v}_n^U}$ of the UAV is defined as ${{( {\boldsymbol{p}_{n + 1}^U - \boldsymbol{p}_n^U})} / \Delta }$, and ${v_{\max}}$ is its maximum velocity. The overall system variables and parameters are summarized in Table I.

\begin{figure}[t!]
    \centering
    \includegraphics{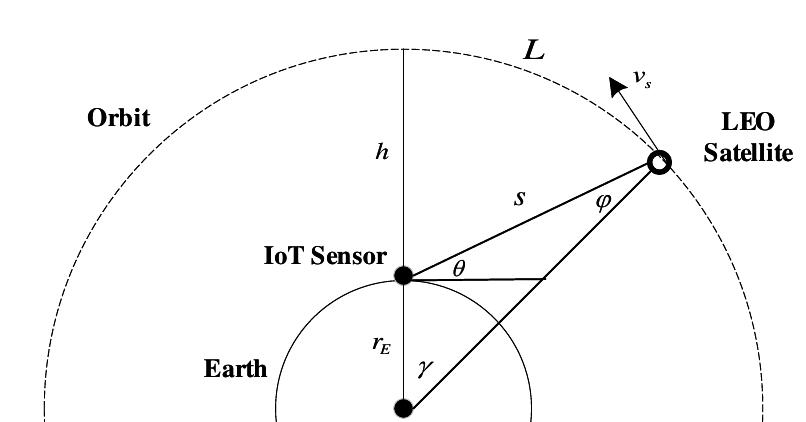}
    \caption{Geometric relationship between the ground user and the LEO satellite.}
\end{figure}

We assume that communication channels between the IoT sensors and UAV [\ref{channel3}], [\ref{channel1}], and between the UAV and LEO satellite [\ref{channel2}], [\ref{channel3}] are dominated by line-of-sight (LoS) links. At the $n$th frame, the channel gains for the IoT sensor $k$-UAV link and UAV-LEO link are written as
\begin{align}
{g_{k,n}}( {\boldsymbol{p}_n^U} ) = \frac{{{g_0}}}{{{{( {x_n^U - x_k^I})}^2} + {{( {y_n^U - y_k^I})}^2}} + {{{h_U}}^2}}
\end{align}
and 
\begin{align}
{h_{n}}( {\boldsymbol{p}_n^U} ) = \frac{{{g_0}{G}}}{{{{( {x_n^L - x_n^U} )}^2} + {{( {y_n^L - y_n^U} )}^2}} + {{{{h_L}}}^2}},
\end{align}
respectively, where ${g_0}$ represents the channel gain at the reference distance 1 m, and $G$ is an antenna gain for the long-distance satellite communication consisting of the transmission antenna gain of the UAV and the receiver antenna gain of the LEO satellite [\ref{channel2}], [\ref{channel4}]. In real applications, note that ${h_n}({\boldsymbol{p}}_n^U) \gg {g_{k,n}}({\boldsymbol{p}}_n^U)$ is guaranteed. For communication links, an additive white Gaussian noise is considered with zero mean and power spectral density ${N_0}$ [dBm/Hz].

\subsection{Coverage Model of the LEO Satellite}

In this section, we explore the beam coverage model [\ref{Sat-Com2}], [\ref{visible1}] of an LEO satellite that accounts for the effect of the orbit of revolution. As shown in Fig. 3, when the LEO satellite makes an orbit round, the available communication time with the UAV can be limited, which is referred to as the LEO visible time window. The length of the visible time window is defined as
\begin{align}
 T_v = \frac{L}{{{v_s}}} = \frac{{2\left( {{r_E} + h} \right)\gamma }}{{{v_s}}}, \label{equ:visible}
\end{align}
where ${v_s}$ is the speed of the LEO satellite. $L$ is the arc length to define the coverage where IoT sensors can communicate with the LEO satellite, and is calculated by $L = {{2\left( {{r_E} + h} \right)\gamma }}$ with ${r_E}$ being the radius of Earth, $h$ being the height of the LEO satellite orbit, and $\gamma$ being the angle of the satellite coverage. In general, due to the very low altitude of a UAV in comparison to the orbit height, the same visible time window is applied to the UAV and IoT sensors. The maximum length of the LEO visible time window can be achieved when ${\gamma  = \pi }$. The angle $\gamma$ of the satellite coverage is calculated by
\begin{align}
 \gamma  = {\cos ^{ - 1}}\left( {\frac{{{r_E}}}{{{r_E} + h}} \cdot \cos \theta } \right) - \theta,
\end{align}
where $\theta$ and $\varphi$ are the elevation angle and the beamwidth of the satellite, respectively, and are derived as $\theta  = {\cos ^{ - 1}}\left( {\frac{{{r_E} + h}}{s} \cdot \cos\left( \theta + \varphi \right)} \right)$ and $\varphi  = {\pi  \mathord{\left/
 {\vphantom {\pi  2}} \right.
 \kern-\nulldelimiterspace} 2} - \left( {\theta  + \gamma } \right)$
 with $s$ indicating the distance between the IoT sensor and LEO satellite. We assume that the UAV can fully access the LEO satellite within the visible time window of ${T_v}$. According to the availability of LEO communication based on the coverage model, three different cases can be considered: ``Always On," ``Always Off" and ``Intermediate Disconnected", the details for which are described below.

\quad \textit{1) ``Always On" scenario ($T \le T_v$)}: The first scenario is when the UAV can communicate with the LEO satellite during the entire mission time since the total mission time is within the LEO visible time, i.e., $T \le T_v$. In this scenario, we have ${\alpha _{k,n}} = 1$ for all $n \in \mathcal{N}$; therefore, the computation capability of the UAV determines whether the UAV or LEO will be used for computing.

\quad \textit{2) ``Always Off" scenario ($T_v = 0$)}: The second scenario is when LEO communication is not available during the entire mission time since the UAV flies outside the beam coverage of the LEO satellite, i.e., $T_v = 0$. In this scenario, we have ${\alpha _{k,n}} = 0$ for all $n \in \mathcal{N}$, and only the UAV computing can be performed. Furthermore, when the offloaded data size exceeds the UAV computation capability, it is transferred to the end user via the UAV without computing.

\begin{table*}[t]
    \caption{Three different scenarios according to LEO availability.}\label{Table_new}
    \centering
    \begin{tabular}{|>{\centering}m{4.5cm}|>{\centering}m{3.7cm}|>{\centering}m{3.7cm}|>{\centering}m{3.5cm}|}
     \hline
     Scenario & $\alpha_{k,n}$ & ${\beta _{k,n}}$ & Available types of computing \tabularnewline
     \hline
     \multirow{4}{3cm}{\centering ``Always On" ($T \le T_v$)} & \multirow{4}{*}{${1, \; {\rm{for}}\; {\rm{all}}\; n \in {\cal N}}$} & \multirow{2}{4cm}{\centering 0, ${\rm{for}}\; {\rm{all}}\;  n \in {\cal N}$} & 
     \multirow{2}{*}{UAV Computing} \tabularnewline
     &&&\tabularnewline\cline{3-3}\cline{4-4}
     && \multirow{2}{4cm}{\centering 1, ${\rm{for}}\; {\rm{all}}\; n \in {\cal N}$} & \multirow{2}{*}{LEO Computing} \tabularnewline
     &&&\tabularnewline
     \hline
     \multirow{2}{3cm}{\centering ``Always Off" ($T_v=0$)} & \multirow{2}{*}{${0, \; {\rm{for}}\;{\rm{all}}\; n \in {\cal N}}$} & \multirow{2}{*}{0, ${\rm{for}}\;{\rm{all}}\; n \in {\cal N}$} & \multirow{2}{*}{UAV Computing} \tabularnewline
     &&&\tabularnewline
     \hline
     \multirow{4}{4.5cm}{\centering ``Intermediate Disconnected" ($T > T_v$)} & \multirow{4}{3.5cm}{\centering $1, \; {\rm{for}}\; n \in \left\{ {1, \cdot  \cdot  \cdot ,N_t} \right\}$, $0, \; {\rm{for}}\; n \in \left\{ {N_t+1, \cdot  \cdot  \cdot ,N} \right\}$} & \multirow{2}{3.7cm}{\centering \; 0, ${\rm{for}}\; n \in \left\{ {1, \cdot  \cdot  \cdot ,N_t} \right\}$, \quad  \quad 0, ${\rm{for}}\; n \in \left\{ {N_t+1, \cdot  \cdot  \cdot ,N} \right\}$} &  \multirow{2}{*}{UAV Computing}\tabularnewline
     &&& \tabularnewline\cline{3-3}\cline{4-4}
     && \multirow{2}{3.7cm}{\centering \; 1, ${\rm{for}}\; n \in \left\{ {1, \cdot  \cdot  \cdot ,N_t} \right\}$, \quad \quad 0, ${\rm{for}}\; n \in \left\{ {N_t+1, \cdot  \cdot  \cdot ,N} \right\}$} &  \multirow{2}{2.5cm}{\centering LEO Computing $\to$ UAV Computing} \tabularnewline
     &&&\tabularnewline
     \hline
    \end{tabular}
\end{table*}

\quad \textit{3) ``Intermediate Disconnected" scenario ($T > T_v$)}: The final scenario is when LEO connection is lost during the mission time, since the total mission time is larger than the LEO visible time, i.e., $T > T_v$. In this scenario, when $t \le T_v$, we have ${\alpha _{k,n}} = 1$ for $n \in \left\{{ 1, \cdot  \cdot  \cdot ,N_t}\right\}$, with $N_t$ being the last frame within $T_v$, where both LEO computing and UAV computing can be performed: that is, $\beta_{k,n} \in \left\{{0, 1} \right\}$. When $t> T_v$, ${\alpha _{k,n}} = 0$ for $n \in \left\{{ N_t+1, \cdot  \cdot  \cdot ,N}\right\}$, where only UAV computing is available: that is, $\beta_{k,n} = 0$. For example, if the LEO connection is lost at $T_v = T/2$, $N_t$ is defined as ${{ {N} } \mathord{\left/ {\vphantom {{ {N} } 2}} \right. \kern-\nulldelimiterspace} 2}$. The frame data of $n \in \left\{{1, \cdot  \cdot  \cdot ,N_t}\right\}$ is computed by the LEO or UAV, while the frame data of $n \in \left\{{ N_t + 1, \cdot  \cdot  \cdot ,N}\right\}$ is computed by the UAV. The details for these three scenarios are summarized in Table II.

\subsection{Energy Consumption Model for Offloading}
In the proposed hierarchical architecture, IoT sensors and the UAV are battery-limited, while the available energy of the LEO satellite is much more sufficient due to its larger size and mass, which is therefore negligible for the system design. With the aim of minimizing the total energy consumption of the UAV, we cover the energy consumption model for computation, communication and flying required for offloading. Here, the LEO satellite is assumed to have sufficient battery capacity compared to the UAV and IoT sensors [\ref{Sat-Com2}], [\ref{Sat-Com1}], which is not reflected in the system design.  

\quad \textit{1) Computation energy model}: First, we define the amount of computation energy consumption at the LEO and UAV-mounted cloudlets at the $n$th frame for IoT sensor $k$ as [\ref{refer_para1}], [\ref{refer_para2}]
\begin{align}
E_{k,n}^d( {l_{k,n}^d}) = \frac{{{\gamma ^d}C_k^dl_{k,n}^d}}{{{\Delta ^2}}}{\left( {\sum\limits_{{k'} = 1}^K {C_{{k'}}^d} l_{{k'},n}^d} \right)^2},\label{equ:comp1}
\end{align}
where $d \in \left\{{L, U}\right\}$ with $L$ indicating the LEO satellite and $U$ indicating the UAV; $l_{k,n}^d$ is the number of bits to be computed at the cloudlet and ${\gamma ^d}$ is the effective switched capacitance of the cloudlet.

\quad \textit{2) Communication energy model}: In the proposed system, the transmit energy consumption from the UAV to LEO at the $n$th frame for offloading the task of the IoT sensor $k$ is defined as [\ref{channel1}], [\ref{refer_SNR}]
\begin{align}
E_{k,n}^{U,L}({L_{k,n}^{U,L},\boldsymbol{p}_n^U}) = \frac{{{N_0}{{B\Delta } \mathord{\left/
 {\vphantom {{B\Delta } K}} \right.
 \kern-\nulldelimiterspace} K}}}{{{h_{n}}({\boldsymbol{p}_n^U})}}\left( {{2^{\frac{{L_{k,n}^{U,L}}}{{{{B\Delta } \mathord{\left/
 {\vphantom {{B\Delta } K}} \right.
 \kern-\nulldelimiterspace} K}}}}} - 1} \right),\label{equ:UAV1}
\end{align}
where $L_{k,n}^{U,L}$ is the number of uplink bits. At the final destination of the UAV above the end user, the downlink communication energy consumption is required so that the UAV can transmit the computing results accumulated during flying, which is given as
\begin{align}
E^{U,E}({L^{U,E},\boldsymbol{p}_{N+1}^U}) = \frac{{{N_0}{{B\Delta } \mathord{\left/
 {\vphantom {{B\Delta } K}} \right.
 \kern-\nulldelimiterspace} K}}}{{{g_{K+1,N+1}}({\boldsymbol{p}_{N+1}^U})}}\left( {{2^{\frac{{L^{U,E}}}{{{{B\Delta } \mathord{\left/
 {\vphantom {{B\Delta } K}} \right.
 \kern-\nulldelimiterspace} K}}}}} - 1} \right),\label{equ:UAV2}
\end{align}
where $L^{U,E}$ is the number of downlink bits and is the same as the sum of output bits of the UAV and LEO-mounted cloudlets as follows: 
\begin{align}
L^{U,E} = O_k^U\sum\limits_{n = 1}^{N - 2} {l_{k,n + 1}^U}  + O_k^L\sum\limits_{n = 1}^{N - 4} {l_{k,n + 2}^L}.
\end{align}
In addition, the transmit energy consumption from the LEO and IoT sensor $k$ 
to the UAV at the $n$th frame is defined as
\begin{align}
E_{k,n}^{L,U}({L_{k,n}^{L,U},\boldsymbol{p}_n^U}) = \frac{{{N_0}{{B\Delta } \mathord{\left/
 {\vphantom {{B\Delta } K}} \right.
 \kern-\nulldelimiterspace} K}}}{{{h_{n}}({\boldsymbol{p}_n^U})}}\left( {{2^{\frac{{L_{k,n}^{L,U}}}{{{{B\Delta } \mathord{\left/
 {\vphantom {{B\Delta } K}} \right.
 \kern-\nulldelimiterspace} K}}}}} - 1} \right)
\end{align}
and
\begin{align}
E_{k,n}^{I,U}({L_{k,n}^{I,U},\boldsymbol{p}_n^U}) = \frac{{{N_0}{{B\Delta } \mathord{\left/
 {\vphantom {{B\Delta } K}} \right.
 \kern-\nulldelimiterspace} K}}}{{{g_{k,n}}({\boldsymbol{p}_n^U})}}\left( {{2^{\frac{{L_{k,n}^{I,U}}}{{{{B\Delta } \mathord{\left/
 {\vphantom {{B\Delta } K}} \right.
 \kern-\nulldelimiterspace} K}}}}} - 1} \right),
 \end{align}
where $L_{k,n}^{L,U}$ is the number of downlink bits transmitted at the LEO and $L_{k,n}^{I,U}$ is the number of uplink bits transmitted at the IoT sensor $k$. The energy consumption for reception is excluded since it is much smaller than the transmission energy consumption.

\quad \textit{3) Flying energy model}: Following [\ref{flying1}], [\ref{flying2}], the flying energy consumption of the UAV at the $n$th frame is written as
\begin{align}
E_{n}^F( {\boldsymbol{v}_n^U}) = \kappa {\| {\boldsymbol{v}_n^U} \|^2},\label{equ:UAV velocity}
\end{align}
where $\kappa  = 0.5M\Delta $ and $M$ is the mass of the UAV. The flying energy consumption depends only on the velocity vector ${\boldsymbol{v}_n^U}$ of the UAV, and the level flight entails no change in the gravitational potential energy.

Our purpose is to minimize the total energy consumption of the UAV, which must be calculated as the sum of the energy consumption of computation, communication and flying: 
\begin{align}
E_{k,n}^{total} & = {\alpha _{k,n}}\left\{ {{\beta _{k,n}}E_{k,n}^{U,L}(L_{k,n}^{U,L},{\boldsymbol{p}}_n^U) + (1 - {\beta _{k,n}})E_{k,n}^U(l_{k,n}^U)} \right\} \nonumber \\
&  \quad  + (1 - {\alpha _{k,n}})(1 - {\beta _{k,n}})E_{k,n}^U(l_{k,n}^U) + E_{n}^F( {\boldsymbol{v}_n^U}),\label{equ:offloading}
\end{align}
where ${\alpha _{k,n}}$ and ${\beta _{k,n}}$ are variables for the LEO availability and scheduling between LEO computing and UAV computing, respectively, which are given as 
\begin{align}
& {{\alpha _{k,n}}} =
\begin{dcases}
{\ 1,}& \ {{\rm{if}\; {\rm{LEO}}\; {\rm{communication}}\; {\rm{is}}\; {\rm{available}}}},\\
{\ 0,}& \ {{\rm{otherwise}}},\\
\end{dcases}\label{equ:alpha}\\
& {{\beta _{k,n}}} =
\begin{dcases}
{\ 1,}& \ {{\rm{if}\; {\rm{LEO}}\; {\rm{computing}}\; {\rm{is}}\; {\rm{performed}}}},\\
{\ 0,}& \ {{\rm{if}\; {\rm{UAV}}\; {\rm{computing}}\; {\rm{is}}\; {\rm{performed}}}}.
\end{dcases}\label{equ:beta}
\end{align}
Note that the energy consumption $E^{U,E}$ for downlink communication with the end user in (\ref{equ:UAV2}) is excluded from (\ref{equ:offloading}) since it is constant regardless of optimization. In addition, LEO computing is considered by $\beta_{k,n}=1$ when the input bits of the IoT sensor $k$ exceeds the computation capability of the UAV: that is,  
\begin{align}
{\sum\limits_{n = 1}^{N} {L_{k,n}^{I,U}} > \sum\limits_{n = 1}^{N} {\left( {f_n^U \cdot \frac{\Delta }{{K}}} \right)\frac{1}{{C_k^U}}}},\label{equ:w2}
\end{align}
where $f_n^U$ [CPU cycles/s] is the CPU frequency at the UAV edge server.

\section{Optimal Energy Consumption for the \\ ``Always On" Scenario}
In this section, we formulate an optimization problem and the proposed algorithm to obtain a solution for the ``Always On" scenario. Depending on the size of the offloaded data, either LEO computing or UAV computing is selected. As mentioned above, the total UAV energy consumption $E_{k,n}^{total}$ in (\ref{equ:offloading}) is rewritten with ${\alpha _{k,n}} = 1$, for all $n \in \mathcal{N}$, as
\begin{align}
E_{k,n}^{total} & = {{\beta _{k,n}}E_{k,n}^{U,L}(L_{k,n}^{U,L},{\boldsymbol{p}}_n^U) + (1 - {\beta _{k,n}})E_{k,n}^U(l_{k,n}^U)} \nonumber \\
& \quad + E_{n}^F( {\boldsymbol{v}_n^U}).\label{sc1_eq}
\end{align}

When LEO computing is considered, i.e., ${\beta _{k,n}}=1$, we need to jointly optimize the bit allocation of ${\{ {L_{k,n}^{I,U}} \}_{n \in \left\{ {1, \cdot  \cdot  \cdot ,N - 4} \right\},k \in \mathcal{K}}}$, ${\{ {L_{k,n}^{U,L}}\}_{n \in \left\{ {2, \cdot  \cdot  \cdot ,N - 3} \right\},k \in \mathcal{K}}}$, ${\{ {l_{k,n}^L} \}_{n \in \left\{ {3, \cdot  \cdot  \cdot ,N - 2} \right\},k \in \mathcal{K}}}$ and ${\{ {L_{k,n}^{L,U}}\}_{n \in \left\{ {4, \cdot  \cdot  \cdot ,N - 1} \right\},k \in \mathcal{K}}}$ along with the UAV trajectory ${\{ {\boldsymbol{p}_n^U}\}_{n \in \left\{ {2, \cdot  \cdot  \cdot ,N} \right\}}}$. When UAV computing is performed, that is, ${\beta _{k,n}}=0$, we must jointly optimize the bit allocation of ${\{ {L_{k,n}^{I,U}}\}_{n \in \left\{ {1, \cdot  \cdot  \cdot ,N - 2} \right\},k \in \mathcal{K}}}$ and ${\{ {l_{k,n}^U} \}_{n \in \left\{ {2, \cdot  \cdot  \cdot ,N - 1} \right\},k \in \mathcal{K}}}$ along with the UAV trajectory ${\{ {\boldsymbol{p}_n^U} \}_{n \in \left\{ {2, \cdot  \cdot  \cdot ,N} \right\}}}$. This problem is formulated with (\ref{sc1_eq}) as follows:
\begin{subequations}
\begin{align}
\mathop {{\rm{min}}}\limits_{L_{k,n}^{I,U},L_{k,n}^{U,L},L_{k,n}^{L,U} \atop
\scriptstyle l_{k,n}^U, l_{k,n}^L,\boldsymbol{p}_n^U} & \sum\limits_{k = 1}^K {\left( {\sum\limits_{n = 1}^{N - 4} {\beta _{k,n}}{E_{k,n + 1}^{U,L}({L_{k,n + 1}^{U,L},\boldsymbol{p}_{n + 1}^U} )} } \right.} \nonumber\\
& \left. { + \sum\limits_{n = 1}^{N - 2} ( {1 - {\beta _{k,n}}}){E_{k,n + 1}^U( {l_{k,n + 1}^U})} } \right) + \sum\limits_{n = 1}^N {E_n^F( {\boldsymbol{v}_n^U})} \label{prob2}
\end{align}
\begin{align}
&{\text{s.t.}} \ {{E_{k,n}^{I,U}(L_{k,n}^{I,U},\boldsymbol{p}_n^U)}}  \le \varepsilon,\; \forall k \in {\cal K}, \; {n \in \mathcal{N}} \label{st1} \\
& \sum\limits_{i = 1}^n {l_{k,i + 1}^U}  \le \sum\limits_{i = 1}^n {L_{k,i}^{I,U}},\; \forall k \in {\cal K},\;n = 1, \cdot  \cdot  \cdot ,N - 2 \label{st4} \\
& \sum\limits_{i = 1}^n {L_{k,i + 1}^{U,L}}  \le \sum\limits_{i = 1}^n {L_{k,i}^{I,U}},\;  \forall k \in {\cal K},\;n = 1, \cdot  \cdot  \cdot ,N - 4 \label{st3} \\
& \sum\limits_{i = 1}^n {l_{k,i + 2}^L}  \le \sum\limits_{i = 1}^n {L_{k,i + 1}^{U,L}},\;  \forall k \in {\cal K},\;n = 1, \cdot  \cdot  \cdot ,N - 4 \label{st5}\\
& \sum\limits_{i = 1}^n {L_{k,i + 3}^{L,U}}  \le O_k^L\sum\limits_{i = 1}^n {l_{k,i + 2}^L},\;  \forall k \in {\cal K},\;n = 1, \cdot  \cdot  \cdot ,N - 4 \label{st6}\\ 
& \sum\limits_{n = 1}^{N - 4} {\beta _{k,n}}{L_{k,n}^{I,U}}  + \sum\limits_{n = 1}^{N - 2} (1 - {\beta _{k,n}}){L_{k,n}^{I,U}}  = {I_k},\;\forall k \in {\cal K} \label{st7}\\
& \sum\limits_{n = 1}^{N - 4} {\beta _{k,n}}{l_{k,n + 2}^L}  + \sum\limits_{n = 1}^{N - 2} (1 - {\beta _{k,n}}){l_{k,n + 1}^U}  = {I_k},\;\forall k \in {\cal K} \label{st8}\\
& \sum\limits_{n = 1}^{N - 4} {l_{k,n + 2}^L} = \sum\limits_{n = 1}^{N - 4} {L_{k,n + 1}^{U,L}},\;\forall k \in {\cal K} \label{st9}\\
& \sum\limits_{n = 1}^{N - 4} {L_{k,n + 3}^{L,U}} = O_k^L\sum\limits_{n = 1}^{N - 4} {L_{k,n + 1}^{U,L}},\;\forall k \in {\cal K} \label{st10}\\
& L_{k,n}^{I,U},L_{k,n}^{U,L},{L_{k,n}^{L,U}},{l_{k,n}^{U}},{l_{k,n}^{L}}\ge 0, \; \forall{k \in \mathcal{K}},\;  {n \in \mathcal{N}}\label{st11}\\
& \boldsymbol{p}_1^U = \boldsymbol{p}_I^U, \ \boldsymbol{p}_{N + 1}^U = \boldsymbol{p}_F^U, \label{st12}\\
& \left\| {\boldsymbol{v}_n^U } \right\| \le {v_{\max }},\ \forall{n \in \mathcal{N}}, \label{st13}
\end{align}
\end{subequations}
where $\varepsilon$ in (\ref{st1}) represents the energy budget constraint per frame for the IoT sensors. The inequality constraint (\ref{st4}) and (\ref{st5}) ensures that the number of bits computed at the UAV and LEO-mounted cloudlet is less than or equal to the number of uplink bits transmitted from the IoT sensor and UAV, respectively. The inequality constraints (\ref{st3}) and (\ref{st6}) ensure that the number of uplink bits from the UAV is less than or equal to the number of uplink bits from the IoT sensor, and the number of downlink bits from the LEO is limited by the number of output bits from the LEO. The equality constraints (\ref{st7}) and (\ref{st8}) enforce that the sum of the uplink bits of the IoT sensor and the sum of the computation bits for the LEO and UAV computing are equal to the input bits of the IoT sensor. The equality constraints (\ref{st9}) and (\ref{st10}) enforce the completion of LEO computing, while (\ref{st11}) is imposed for the non-negative bit allocations. The constraints (\ref{st12}) and (\ref{st13}) represent the flying UAV's initial and final position constraint and the maximum speed constraint, respectively.

Problem (18) is non-convex because the objective function and the energy budget constraint are non-convex. To address this non-convexity, we apply the SCA-based strategy [\ref{SCA1}], [\ref{SCA2}] which builds on the inner convex approximation framework. In particular, we develop proposed algorithm 1 by using the following lemmas.

\quad \textit{Lemma 1}: Given that a non-convex objective function $U(\boldsymbol{x})={f_1}(\boldsymbol{x}){f_2}(\boldsymbol{x})$ is the product of $f_1$ and $f_2$ convex and non-negative for any $\boldsymbol{y}$ in the domain of $U(\boldsymbol{x})$, a convex approximation that satisfies the conditions required by the SCA algorithm is given as
\begin{align}
\bar U\left( {\boldsymbol{x};\boldsymbol{y}} \right) & = {f_1}(\boldsymbol{x}){f_2}(\boldsymbol{y}) + {f_1}(\boldsymbol{y}){f_2}(\boldsymbol{x}) \nonumber \\
& \quad + \frac{{{\tau _i}}}{2}{(\boldsymbol{x} - \boldsymbol{y})^{\rm{T}}}\boldsymbol{H}(\boldsymbol{y})(\boldsymbol{x} - \boldsymbol{y}),
\end{align}
where $\tau _i>0$ is a positive constant, $\boldsymbol{H}(\boldsymbol{y})$ is a positive definite matrix, and ${\left(  \cdot  \right)^{\rm{T}}}$ indicates the transpose.

\quad \textit{Lemma 2}: Given a non-convex constraint $g(\boldsymbol{x}_1,\boldsymbol{x}_2) \le 0$, where $g(\boldsymbol{x}_1,\boldsymbol{x}_2)=h_1(\boldsymbol{x}_1)h_2(\boldsymbol{x}_2)$ is the product of the $h_1$ and $h_2$ convex and non-negative, for any $(\boldsymbol{y}_1,\boldsymbol{y}_2)$ in the domain of $g(\boldsymbol{x}_1,\boldsymbol{x}_2)$, a convex approximation that satisfies the conditions required by the SCA algorithm is given as
\begin{flalign*}
\ \ \ \bar g\left( {{\boldsymbol{x}_1},{\boldsymbol{x}_2};{\boldsymbol{y}_1},{\boldsymbol{y}_2}} \right) &&
\end{flalign*}
\vspace{-0.5cm}
\begin{flalign}
& \ \ \buildrel  \Delta \over = \frac{1}{2}{( {{h_1}({\boldsymbol{x}_1}) + {h_2}({\boldsymbol{x}_2})} )^2} - \frac{1}{2}( {{h_1}^2({\boldsymbol{y}_1}) + {h_2}^2({\boldsymbol{y}_2})} ) \nonumber\\
& \ \ - {h_1}({\boldsymbol{y}_1}){h_1}^{'} ({\boldsymbol{y}_1})({\boldsymbol{x}_1} - {\boldsymbol{y}_1}) - {h_2}({\boldsymbol{y}_2}){h_2}^{'} ({\boldsymbol{y}_2})({\boldsymbol{x}_2} - {\boldsymbol{y}_2}), &&
\end{flalign}
where the partial derivative of $f\left(  \cdot  \right)$ is ${f^{'}}\left(  \cdot  \right)$.

We set the primal variables for the formulated Problem (18) as $\boldsymbol{z}$ $=$ ${\{ {\boldsymbol{z}_n}\} _{n \in \mathcal{N}}}$ with ${\boldsymbol{z}_n}$ $=$ $({\{ L_{k,n}^{I,U}\} _{k \in \mathcal{K}}}$, ${\{ L_{k,n}^{U,L}\} _{k \in \mathcal{K}}}$, ${\{ L_{k,n}^{L,U}\} _{k \in \mathcal{K}}}$, ${\{ l_{k,n}^U\} _{k \in \mathcal{K}}}$, ${\{ l_{k,n}^L\} _{k \in \mathcal{K}}}$, $\boldsymbol{p}_n^U)$. We observe that the function $E_{k,n}^{U,L}({\boldsymbol{z}_n}) \buildrel \Delta \over = E_{k,n}^{U,L}(L_{k,n}^{U,L},\boldsymbol{p}_n^U)$ in (\ref{prob2}) is the product of two convex and non-negative functions, namely
\begin{align}
f_1(L_{k,n}^{U,L})=\frac{{{N_0}B{\Delta  \mathord{\left/
 {\vphantom {\Delta  K}} \right.
 \kern-\nulldelimiterspace} K}}}{{g_0}{G}}\left( {{2^{\frac{{L_{k,n}^{U,L}}}{{B{\Delta  \mathord{\left/
 {\vphantom {\Delta  K}} \right.
 \kern-\nulldelimiterspace} K}}}}} - 1} \right)
\end{align}
and
\begin{align}
f_2(\boldsymbol{p}_{n}^{U})={(x_n^L - x_n^U)^2} + {(y_n^L - y_n^U)^2} + {h_L}^2.
\end{align}
Then, by using Lemma 1 and defining ${\boldsymbol{z}_n(v)} =$ $({\{ L_{k,n}^{I,U}(v)\} _{k \in \mathcal{K}}}$, ${\{ L_{k,n}^{U,L}(v)\} _{k \in \mathcal{K}}}$, ${\{ L_{k,n}^{L,U}(v)\} _{k \in \mathcal{K}}}$, ${\{ l_{k,n}^U(v)\} _{k \in \mathcal{K}}}$, ${\{ l_{k,n}^L(v)\} _{k \in \mathcal{K}}}$, $\boldsymbol{p}_n^U(v)){\in \mathcal{X}}$ for the $v$th iterate within the feasible set $\mathcal{X}$ of (18), we obtain a strongly convex surrogate function $\bar E_{k,n}^{U,L}({\boldsymbol{z}_n};{\boldsymbol{z}_n(v)})$ of $E_{k,n}^{U,L}({\boldsymbol{z}_n})$ as
\begin{flalign*}
\quad \ \ \bar E_{k,n}^{U,L}({\boldsymbol{z}_n};{\boldsymbol{z}_n}(v)) \buildrel \Delta \over = \bar E_{k,n}^{U,L}(L_{k,n}^{U,L},\boldsymbol{p}_n^U;L_{k,n}^{U,L}(v),\boldsymbol{p}_n^U(v))&&
\end{flalign*}
\vspace{-0.5cm}
\begin{flalign}
& = {f_1}(L_{k,n}^{U,L}){f_2}(\boldsymbol{p}_n^U(v)) + {f_1}(L_{k,n}^{U,L}(v)){f_2}(\boldsymbol{p}_n^U) \nonumber\\
& + \frac{{{\tau _{L_{k,n}^{U,L}}}}}{2}{(L_{k,n}^{U,L} - L_{k,n}^{U,L}(v))^2} + \frac{{{\tau _{x_n^U}}}}{2}{(x_n^U - x_n^U(v))^2}\nonumber\\
& + \frac{{{\tau _{y_n^U}}}}{2}{(y_n^U - y_n^U(v))^2}, \label{sol1}
\end{flalign}
where ${\tau _{L_{k,n}^{U,L}}},{\tau _{x_n^U}},{\tau _{y_n^U}}>0$. Also, the function $E_{k,n}^U({\boldsymbol{z}_n}) \buildrel \Delta \over = E_{k,n}^U(l_{k,n}^U)$ in (18a) is the product of two convex and non-negative functions, namely
\begin{align}
{f_1}(l_{k,n}^U) = \frac{{{\gamma ^U}C_k^Ul_{k,n}^U}}{{{\Delta ^2}}}
\end{align}
and
\begin{align}
{f_2}(l_{k',n}^U) = {\left( {\sum\limits_{k' = 1}^K {C_{k'}^U} l_{k',n}^U} \right)^2}.
\end{align}
As in (\ref{sol1}), we obtain a strongly convex surrogate function $\bar E_{k,n}^U({\boldsymbol{z}_n};{\boldsymbol{z}_n(v)})$ of $E_{k,n}^U({\boldsymbol{z}_n})$ as 
\begin{flalign*}
\quad \ \ \bar E_{k,n}^U({\boldsymbol{z}_n};{\boldsymbol{z}_n}(v)) \buildrel \Delta \over = \bar E_{k,n}^U(l_{k,n}^U,l_{k',n}^U;l_{k,n}^U(v),l_{k',n}^U(v))&&
\end{flalign*}
\vspace{-0.5cm}
\begin{flalign}
& = {f_1}(l_{k,n}^U){f_2}(l_{k',n}^U(v)) + {f_1}(l_{k,n}^U(v)){f_2}(l_{k',n}^U)\nonumber\\
& + \frac{{\tau _{l_{k,n}^U}}}{2}{(l_{k,n}^U - l_{k,n}^U(v))^2} + \frac{{\tau _{l_{k',n}^U}}}{2}{(l_{k',n}^U - l_{k',n}^U(v))^2}, \label{sol2}
\end{flalign}
where ${\tau _{l_{k,n}^U}},{\tau _{l_{k',n}^U}}>0$.

For the non-convex energy budget constraint (\ref{st1}), we derive a convex upper bound by using Lemma 2. The function $E_{k,n}^{I,U}({\boldsymbol{z}_n}) \buildrel \Delta \over = E_{k,n}^{I,U}(L_{k,n}^{I,U},\boldsymbol{p}_n^U)$ is the product of two convex and non-negative functions, namely
\begin{align}
{h_1}(L_{k,n}^{I,U}) = {2^{\frac{{L_{k,n}^{I,U}}}{{{{B\Delta } \mathord{\left/
 {\vphantom {{B\Delta } K}} \right.
 \kern-\nulldelimiterspace} K}}}}} - 1
\end{align}
and
\begin{align}
{h_2}(\boldsymbol{p}_n^U) = {(x_n^U - x_k^I)^2} + {(y_n^U - y_k^I)^2} + {h_U}^2.
\end{align}
Then, by using Lemma 2 and defining ${\boldsymbol{z}_n(v)}$ for the $v$th iterate, we obtain a strongly convex surrogate function $\bar E_{k,n}^{I,U}({\boldsymbol{z}_n};{\boldsymbol{z}_n(v)})$ of $E_{k,n}^{I,U}({\boldsymbol{z}_n})$ as
\begin{flalign*}
\ \bar E_{k,n}^{I,U}({\boldsymbol{z}_n};{\boldsymbol{z}_n}(v)) \buildrel \Delta \over = E_{k,n}^{I,U}(L_{k,n}^{I,U},\boldsymbol{p}_n^U;L_{k,n}^{I,U}(v),\boldsymbol{p}_n^U(v))&&
\end{flalign*}
\vspace{-0.5cm}
\begin{flalign}
& = \frac{{{{{N_0}B\Delta } \mathord{\left/
 {\vphantom {{{N_0}B\Delta } K}} \right.
 \kern-\nulldelimiterspace} K}}}{{2{g_0}}}\left[ {{{\left( {{2^{\frac{{L_{k,n}^{I,U}}}{{{{B\Delta } \mathord{\left/
 {\vphantom {{B\Delta } K}} \right.
 \kern-\nulldelimiterspace} K}}}}} - 1 + {{(x_n^U - x_k^I)}^2} + {{(y_n^U - y_k^I)}^2} + {h_U}^2} \right)}^2}} \right.\nonumber \\
& \left. { - {{\left( {{2^{\frac{{L_{k,n}^{I,U}(v)}}{{{{B\Delta } \mathord{\left/
 {\vphantom {{B\Delta } K}} \right.
 \kern-\nulldelimiterspace} K}}}}} - 1} \right)}^2} - {{\left( {{{(x_n^U(v) - x_k^I)}^2} + {{(y_n^U(v) - y_k^I)}^2} + {h_U}^2} \right)}^2}} \right]\nonumber \\
&  - \frac{{{N_0}\ln 2}}{{{g_0}}}{2^{\frac{{L_{k,n}^{I,U}(v)}}{{{{B\Delta } \mathord{\left/
 {\vphantom {{B\Delta } K}} \right.
 \kern-\nulldelimiterspace} K}}}}}\left( {{2^{\frac{{L_{k,n}^{I,U}(v)}}{{{{B\Delta } \mathord{\left/
 {\vphantom {{B\Delta } K}} \right.
 \kern-\nulldelimiterspace} K}}}}} - 1} \right)\left( {L_{k,n}^{I,U} - L_{k,n}^{I,U}(v)} \right) \nonumber \\
& - \frac{{{{2{N_0}B\Delta } \mathord{\left/
 {\vphantom {{2{N_0}B\Delta } K}} \right.
 \kern-\nulldelimiterspace} K}}}{{{g_0}}}\left( {{{(x_n^U(v) - x_k^I)}^2} + {{(y_n^U(v) - y_k^I)}^2} + {h_U}^2} \right) \nonumber \\
 & \left( {(x_n^U(v) - x_k^I)(x_n^U - x_n^U(v)) + (y_n^U(v) - y_k^I)(y_n^U - y_n^U(v))} \right).
\end{flalign}

\begin{algorithm}[t]
\begin{algorithmic}[1]
\caption{Proposed algorithm for the ``Always On" scenario} \label{algorithm_joint}
    \REQUIRE $\gamma (v) \in (0,1]$,  $\boldsymbol{z}(0) = {\{ {\boldsymbol{z}_n}(0)\} _{n \in \mathcal{N}}} \in \mathcal{X}$;   
    Set $v=0$.
    \ENSURE $\{L_{k,n}^{I,U}\},\{L_{k,n}^{U,L}\},\{L_{k,n}^{L,U}\},\{l_{k,n}^U\}, \{l_{k,n}^L\},\{\boldsymbol{p}_n^U\}$.
        \STATE If $\boldsymbol{z}(v)$ is a stationary solution of (18): STOP. 
        \STATE Compute $\hat {\boldsymbol{z}}\left( {\boldsymbol{z}(v)}\right)$ of (30) using dual decomposition or CVX.
        \STATE Set $\boldsymbol{z}(v + 1) = \boldsymbol{z}(v) + \gamma (v)\left( {\hat {\boldsymbol{z}}\left( {\boldsymbol{z}(v)} \right) - \boldsymbol{z}(v)} \right)$.
        \STATE $v \leftarrow v + 1$ and go to step 1.
\end{algorithmic}
\end{algorithm}

Finally, the problem in Equation (18) can be transformed into the strongly convex inner approximation for a given feasible $\boldsymbol{z}(v) = {\{ {\boldsymbol{z}_n(v)}\} _{n \in \mathcal{N}}}$, as
\begin{subequations}
\begin{align}
\mathop {{\rm{min}}}\limits_{\boldsymbol{z}} & \sum\limits_{k = 1}^K {\left( {\sum\limits_{n = 1}^{N - 4} {\beta _{k,n}}{\bar E_{k,n + 1}^{U,L}({\boldsymbol{z}_{n + 1}};{\boldsymbol{z}_{n + 1}}(v))} } \right.} \nonumber\\
& \left. {+\sum\limits_{n = 1}^{N - 2} (1 - {\beta _{k,n}}){\bar E_{k,n + 1}^U({\boldsymbol{z}_{n + 1}};{\boldsymbol{z}_{n + 1}}(v))} } \right) + \sum\limits_{n = 1}^N {E_n^F(\boldsymbol{v}_n^U)} \label{newprob2}
\end{align}
\begin{align}
{\text{s.t.}} \ & {{\bar E_{k,n}^{I,U}({\boldsymbol{z}_n};{\boldsymbol{z}_n}(v))} } \le \varepsilon,\; \forall k \in {\cal K},\; {n \in \mathcal{N}} \label{newst1} \\
& \rm{(\ref{st4})} - \rm{(\ref{st13})}, \label{newst3}
\end{align}
\end{subequations}
which has a unique solution denoted by $\hat {\boldsymbol{z}}\left( {\boldsymbol{z}(v)} \right)$. Since Problem (30) is convex, we can obtain the closed-form solutions via dual decomposition [\ref{CVX1}] or a standard convex optimization solver such as CVX [\ref{CVX2}]. The proposed algorithm based on the SCA method is summarized as Algorithm 1. The sequence $\{\boldsymbol{z}(v)\}$ generated by Algorithm 1 converges if the step size $\gamma (v)$ is chosen so that $\gamma (v) \in (0,1]$, $\gamma (v) \to 0$, and $\sum\nolimits_v {\gamma (v)}  = \infty$. Also, $\{\boldsymbol{z}(v)\}$ is bounded and every limit point of $\{\boldsymbol{z}(v)\}$ is stationary. Furthermore, if Algorithm 1 does not stop after a finite number of steps, none of the stationary points are a local minimum of Problem (18).

\section{Optimal Energy Consumption for the \\ ``Always Off" Scenario}
In this section, we find the optimal bit allocation and UAV path planning when the LEO communication is not available during the entire mission time. Therefore, the total UAV energy consumption $E_{k,n}^{total}$ in (\ref{equ:offloading}) is rewritten with ${\alpha _{k,n}} = 0$ for all $n \in \mathcal{N}$, as
\begin{align}
E_{k,n}^{total} = (1 - {\beta _{k,n}})E_{k,n}^U(l_{k,n}^U) + E_{n}^F( {\boldsymbol{v}_n^U}).\label{sc2_eq}
\end{align}
For UAV computing with ${\beta _{k,n}}=0$, the problem is given with (\ref{sc2_eq}) by
\begin{subequations}
\begin{align}
\mathop {{\rm{min}}}\limits_{L_{k,n}^{I,U},l_{k,n}^U,\scriptstyle \boldsymbol{p}_n^U} & \sum\limits_{k = 1}^K  \sum\limits_{n = 1}^{N - 2} ( {1 - {\beta _{k,n}}}){E_{k,n + 1}^U( {l_{k,n + 1}^U})}  + \sum\limits_{n = 1}^N {E_n^F( {\boldsymbol{v}_n^U})} 
\end{align}
\begin{align}
& {\text{s.t.}} \ \sum\limits_{n = 1}^{N - 2} (1 - {\beta _{k,n}}){L_{k,n}^{I,U}}  = {I_k},\;\forall k \in {\cal K}\label{st1_sc2}\\
& \sum\limits_{n = 1}^{N - 2} (1 - {\beta _{k,n}}){l_{k,n + 1}^U}  = {I_k},\;\forall k \in {\cal K}\label{st2_sc2}\\
& \rm{(\ref{st1})}, \rm{(\ref{st4})}, \rm{(\ref{st11})} - \rm{(\ref{st13})},\label{st3_sc2}
\end{align}
\end{subequations}
where the equality constraints (\ref{st1_sc2}) and (\ref{st2_sc2}) guarantee that the total number of uplink bits from the IoT sensor and the total number of computation bits at the UAV must be equal to the input bits of the IoT sensor for complete offloading.

\begin{algorithm}[t]
\begin{algorithmic}[1]
\caption{Proposed algorithm for the ``Always Off" scenario} \label{algorithm_joint1}
    \REQUIRE $\gamma (v) \in (0,1]$,  $\boldsymbol{z}(0) = {\{ {\boldsymbol{z}_n}(0)\} _{n \in \mathcal{N}}} \in \mathcal{X}$; Set $v=0$.
    \ENSURE $\{L_{k,n}^{I,U}\},\{l_{k,n}^U\},\{\boldsymbol{p}_n^U\}$.
        \STATE If $\boldsymbol{z}(v)$ is a stationary solution of (32): STOP.
        \STATE Compute $\hat {\boldsymbol{z}}\left( {\boldsymbol{z}(v)}\right)$ of (33) using dual decomposition or CVX.
        \STATE Set $\boldsymbol{z}(v + 1) = \boldsymbol{z}(v) + \gamma (v)\left( {\hat {\boldsymbol{z}}\left( {\boldsymbol{z}(v)} \right) - \boldsymbol{z}(v)} \right)$.
        \STATE $v \leftarrow v + 1$ and go to step 1.
\end{algorithmic}
\end{algorithm}

In the ``Always Off" case, the primal variables are defined as $\boldsymbol{z} = {\{ {\boldsymbol{z}_n}\} _{n \in \mathcal{N}}}$ with ${\boldsymbol{z}_n} = ({\{ L_{k,n}^{I,U}\} _{k \in \mathcal{K}}}$, ${\{ l_{k,n}^U\} _{k \in \mathcal{K}}}$, $\boldsymbol{p}_n^U)$. Since Problem (32) is non-convex, it can be transformed into the strongly convex inner approximation, for a given a feasible $\boldsymbol{z}(v) = {\{ {\boldsymbol{z}_n(v)}\} _{n \in \mathcal{N}}}$, as
\begin{subequations}
\begin{align}
\mathop {{\rm{min}}}\limits_{\boldsymbol{z}} & \sum\limits_{k = 1}^K {\sum\limits_{n = 1}^{N - 2} (1 - {\beta _{k,n}}){\bar E_{k,n + 1}^U({\boldsymbol{z}_{n + 1}};{\boldsymbol{z}_{n + 1}}(v))} } + \sum\limits_{n = 1}^N {E_n^F(\boldsymbol{v}_n^U)} \label{newprob2_sc2}
\end{align}
\begin{align}
{\text{s.t.}} \ & \rm{(\ref{st1_sc2})}, \rm{(\ref{st2_sc2})}, \rm{(\ref{newst1})}, \rm{(\ref{st4})}, \rm{(\ref{st11})} - \rm{(\ref{st13})}, \label{newst3_sc2}
\end{align}
\end{subequations}
where $\bar E_{k,n}^U$ of the objective function is defined equally in (\ref{sol2}). Problem (33) has a unique solution denoted by $\hat {\boldsymbol{z}}\left( {\boldsymbol{z}(v)} \right)$ due to its convexity. As in Problem (30), the locally optimal solution can be obtained by dual decomposition or a standard convex optimization solver. The proposed SCA-based algorithm is summarized in Algorithm 2.

\begin{algorithm}[t]
\begin{algorithmic}[1]
\caption{Proposed algorithm for the ``Intermediate Disconnected" scenario} \label{algorithm_joint2}
    \REQUIRE $\gamma (v) \in (0,1]$,  $\boldsymbol{z}(0) = {\{ {\boldsymbol{z}_n}(0)\} _{n \in \mathcal{N}}} \in \mathcal{X}$; Set $v=0$.
    \ENSURE $\{L_{k,n}^{I,U}\},\{L_{k,n}^{U,L}\},\{L_{k,n}^{L,U}\},\{l_{k,n}^U\}, \{l_{k,n}^L\},\{\boldsymbol{p}_n^U\}$.
        \STATE If $\boldsymbol{z}(v)$ is a stationary solution of (34): STOP. 
        \STATE Compute $\hat {\boldsymbol{z}}\left( {\boldsymbol{z}(v)} \right)$ of (35) using dual decomposition or CVX.
        \STATE Set $\boldsymbol{z}(v + 1) = \boldsymbol{z}(v) + \gamma (v)\left( {\hat {\boldsymbol{z}}\left( {\boldsymbol{z}(v)} \right) - \boldsymbol{z}(v)} \right)$.
        \STATE $v \leftarrow v + 1$ and go to step 1.
\end{algorithmic}
\end{algorithm}

\section{Optimal Energy Consumption for the ``Intermediate Disconnected" Scenario}
For the ``Intermediate Disconnected" case, we provide joint path planning and resource allocation when the LEO communication is intermediately disconnected. The total UAV energy consumption in this case follows (\ref{equ:offloading}).

During the LEO computing for $n \in \left\{{1, \cdot  \cdot  \cdot,  N_t}\right\}$ with ${\alpha _{k,n}}=1$ and ${\beta _{k,n}}=1$, we jointly optimize the bit allocation ${\{ {L_{k,n}^{I,U}} \}_{n \in \left\{ {1, \cdot  \cdot  \cdot ,N_t} \right\},k \in \mathcal{K}}}$, ${\{ {L_{k,n}^{U,L}}\}_{n \in \left\{ {2, \cdot  \cdot  \cdot ,N_t+1} \right\},k \in \mathcal{K}}}$, ${\{ {l_{k,n}^L} \}_{n \in \left\{ {3, \cdot  \cdot  \cdot ,N_t+2} \right\},k \in \mathcal{K}}}$ and ${\{ {L_{k,n}^{L,U}}\}_{n \in \left\{ {4, \cdot  \cdot  \cdot ,N_t+3} \right\},k \in \mathcal{K}}}$ along with the UAV trajectory ${\{ {\boldsymbol{p}_n^U}\}_{n \in \left\{ {2, \cdot  \cdot  \cdot ,N_t+4} \right\}}}$. During UAV computing for $n \in \left\{{1, \cdot  \cdot  \cdot, N_t}\right\}$ with ${\alpha _{k,n}}=1$ and ${\beta _{k,n}}=0$ and $n \in \left\{{N_t+1, \cdot  \cdot  \cdot, N}\right\}$ with ${\alpha _{k,n}}=0$ and ${\beta _{k,n}}=0$, the bit allocation and the UAV path planning are jointly designed as in the UAV computing process of the ``Always On" case. Accordingly, we can formulate the problem as 
\begin{subequations}
\begin{align}
\mathop {{\rm{min}}}\limits_{L_{k,n}^{I,U},L_{k,n}^{U,L},L_{k,n}^{L,U} \atop
\scriptstyle l_{k,n}^U, l_{k,n}^L,\boldsymbol{p}_n^U} & \sum\limits_{k = 1}^K  { \sum\limits_{n = 1}^{N_t} {\alpha _{k,n}} \left\{ {\beta _{k,n}}{E_{k,n + 1}^{U,L}({L_{k,n + 1}^{U,L},\boldsymbol{p}_{n + 1}^U} )}  \right.} \nonumber\\
& \left. + \left( {1 - {{\beta }_{k,n}}} \right){E_{k,n + 1}^U(l_{k,n + 1}^{U})}\right\} \nonumber\\
&  + \sum\limits_{k = 1}^K\sum\limits_{n = N_t+1}^{N - 2} \left( {1 - {{\alpha }_{k,n}}} \right)( {1 - {\beta _{k,n}}}){E_{k,n + 1}^U( {l_{k,n + 1}^U})}\nonumber\\
& + \sum\limits_{n = 1}^N {E_n^F( {\boldsymbol{v}_n^U})} \label{prob2_sc3}
\end{align}
\begin{align}
&{\text{s.t.}} \ \sum\limits_{i = 1}^n {L_{k,i + 1}^{U,L}}  \le \sum\limits_{i = 1}^n {L_{k,i}^{I,U}},\; \forall k \in {\cal K}, \;n = 1, \cdot  \cdot  \cdot ,N_t \label{st2_sc3} \\
& \sum\limits_{i = 1}^n {l_{k,i + 2}^L}  \le \sum\limits_{i = 1}^n {L_{k,i + 1}^{U,L}},\; \forall k \in {\cal K}, \;n = 1, \cdot  \cdot  \cdot ,N_t \label{st5_sc3}\\
& \sum\limits_{i = 1}^n {L_{k,i + 3}^{L,U}}  \le O_k^L\sum\limits_{i = 1}^n {l_{k,i + 2}^L},\; \forall k \in {\cal K}, \;n = 1, \cdot  \cdot  \cdot ,N_t \label{st6_sc3}\\ 
& \sum\limits_{n = 1}^{N_t} {\beta _{k,n}}{l_{k,n + 2}^L} + \sum\limits_{n = 1}^{N - 2}(1 - {\beta _{k,n}}){l_{k,n+1}^{U}}  = {I_k},\;\forall k \in {\cal K} \label{st8_sc3}\\
& \sum\limits_{n = 1}^{N_t} {l_{k,n + 2}^L} = \sum\limits_{n = 1}^{N_t} {L_{k,n + 1}^{U,L}},\;\forall k \in {\cal K} \label{st9_sc3}\\
& \sum\limits_{n = 1}^{N_t} {L_{k,n + 3}^{L,U}} = O_k^L\sum\limits_{n = 1}^{N_t} {L_{k,n + 1}^{U,L}} ,\;\forall k \in {\cal K} \label{st10_sc3}\\
& \rm{(\ref{st1})}, \rm{(\ref{st4})}, \rm{(\ref{st1_sc2})}, \rm{(\ref{st11})} - \rm{(\ref{st13})}, \label{st11_sc3}
\end{align}
\end{subequations}
where the inequality constraints (\ref{st2_sc3})-(\ref{st6_sc3}) and equality constraints (\ref{st8_sc3})-(\ref{st10_sc3}) limit the number of frames to $n = 1, \cdot  \cdot  \cdot ,N_t$ instead of $n = 1, \cdot  \cdot  \cdot ,N - 4$ in constraints (\ref{st3})-(\ref{st6}) and (\ref{st8})-(\ref{st10}), respectively.

In the``Intermediate Disconnected" case, the primal variables are defined the same as in the``Always On" case. By applying the SCA method,the non-convex Problem (34) can be transformed into the strongly convex inner approximation for a given a feasible $\boldsymbol{z}(v) = {\{ {\boldsymbol{z}_n(v)}\} _{n \in \mathcal{N}}}$, as
\begin{subequations}
\begin{align}
\mathop {{\rm{min}}}\limits_{\boldsymbol{z}}&\sum\limits_{k = 1}^K  { \sum\limits_{n = 1}^{N_t} {\alpha _{k,n}} \left\{ {\beta _{k,n}}{\bar E_{k,n + 1}^{U,L}({\boldsymbol{z}_{n + 1}};{\boldsymbol{z}_{n + 1}}(v))}  \right.} \nonumber\\
&\left.+\left( {1 - {{\beta }_{k,n}}} \right){\bar E_{k,n + 1}^U({\boldsymbol{z}_{n + 1}};{\boldsymbol{z}_{n + 1}}(v))}\right\} \nonumber\\
&+\sum\limits_{k = 1}^K\sum\limits_{n = N_t+1}^{N - 2} \left( {1 - {{\alpha }_{k,n}}} \right)( {1 - {\beta _{k,n}}}){\bar E_{k,n + 1}^U({\boldsymbol{z}_{n + 1}};{\boldsymbol{z}_{n + 1}}(v))}\nonumber\\
&+\sum\limits_{n = 1}^N {E_n^F( {\boldsymbol{v}_n^U})} \label{newprob2_sc3}
\end{align}
\begin{align}
{\text{s.t.}} \ & \rm{(\ref{st2_sc3})} - \rm{(\ref{st10_sc3})}, \rm{(\ref{newst1})}, \rm{(\ref{st4})}, \rm{(\ref{st1_sc2})}, \rm{(\ref{st11})} - \rm{(\ref{st13})}, \label{newst3_sc3}
\end{align}
\end{subequations}
which has a unique solution denoted by $\hat {\boldsymbol{z}}\left( {\boldsymbol{z}(v)} \right)$ to be obtained by dual decomposition or a standard convex optimization solver. Algorithm 3 describes the proposed method for the ``Intermediate Disconnected" scenario.

\begin{table}[t!]
    \caption{Simulation Parameters}\label{Table1}
    \centering
    \begin{tabular}{ | c | c || c | c|  }
     \hline
     Parameter&Value&Parameter&Value\\
     \hline
     $v_s$ & 7.5 km/s & $r_E$ & 6371 km\\
     \hline
     $\theta$ & 10 $^{\circ}$ & $T_v$ & 830 s \\
     \hline
     $h_U$ & 1 km & $h_L$ & 600 km\\
     \hline
     $K$ & 10 & $v_{\max }$ & 50 m/s\\
     \hline
     \rule{0pt}{8pt} $M$ & 9.65 kg & $O_k^L$, $O_k^U$ & 0.5\\
     \hline
     \rule{0pt}{8pt} ${f_n^U}$ & $19.5 \times {10^9}$ cycles/s [\ref{Sat-Com2}] & $G$ & $10$ dB\\
     \hline
     \rule{0pt}{8pt} ${\gamma ^L}$, ${\gamma ^U}$ & ${10^{-28}}$ [\ref{refer_para1}], [\ref{refer_para2}] & ${C_k^L}$, ${C_k^U}$ & 1550.7 [\ref{refer_para1}], [\ref{refer_para2}] \\    
     \hline
     $B$ & 40 MHz & ${N_0}$ & -174 dBm/Hz \\  
     \hline
     $\varepsilon$ & 0.11 J & ref. SNR & 80 dB \\ 
     \hline
    \end{tabular}
\end{table}

\section{Simulation Results}                    
In this section, we evaluate the performance of the proposed algorithms to jointly optimize the bit allocation and the UAV trajectory for marine IoT systems in various LEO accessible statuses. For reference, we consider the following schemes: \textit{(i) No optimization}: The equal bit allocation is considered for communication and computation per frame, while the UAV flies at constant velocity between the initial and final positions as $\boldsymbol{p}_n^U = \boldsymbol{p}_I^U + {{\left( {n - 1} \right)\left( {\boldsymbol{p}_F^U - \boldsymbol{p}_I^U} \right)} \mathord{\left/
 {\vphantom {{\left( {n - 1} \right)\left( {\boldsymbol{p}_F^U - \boldsymbol{p}_I^U} \right)} N}} \right.
 \kern-\nulldelimiterspace} N}$, ${\rm{for}} \; n \in \mathcal{N}$; \textit{(ii) Optimized bit allocation}: The communication and computation bits are optimized by the proposed algorithms while considering the UAV trajectory with the constant-velocity as in \textit{(i)}; \textit{(iii) Optimized UAV trajectory}: The path planning of the UAV is obtained by the proposed algorithms with fixed equal bit allocation per frame. The simulation parameters are provided in Table III. Particularly, the space segment considers Iridium-like LEO satellite networks that provide global coverage with 66 satellites distributed in 6 polar orbits [\ref{channel2}], where the orbit height is $h = $ 601 km with the elevation angle $ \theta = $ 10 $^{\circ}$, and satellites in the orbit travel at a speed of around $v_s = $ 7.5 km/s.

 \begin{figure}[t]
     \centering
         \centering
         \includegraphics[width=\columnwidth]{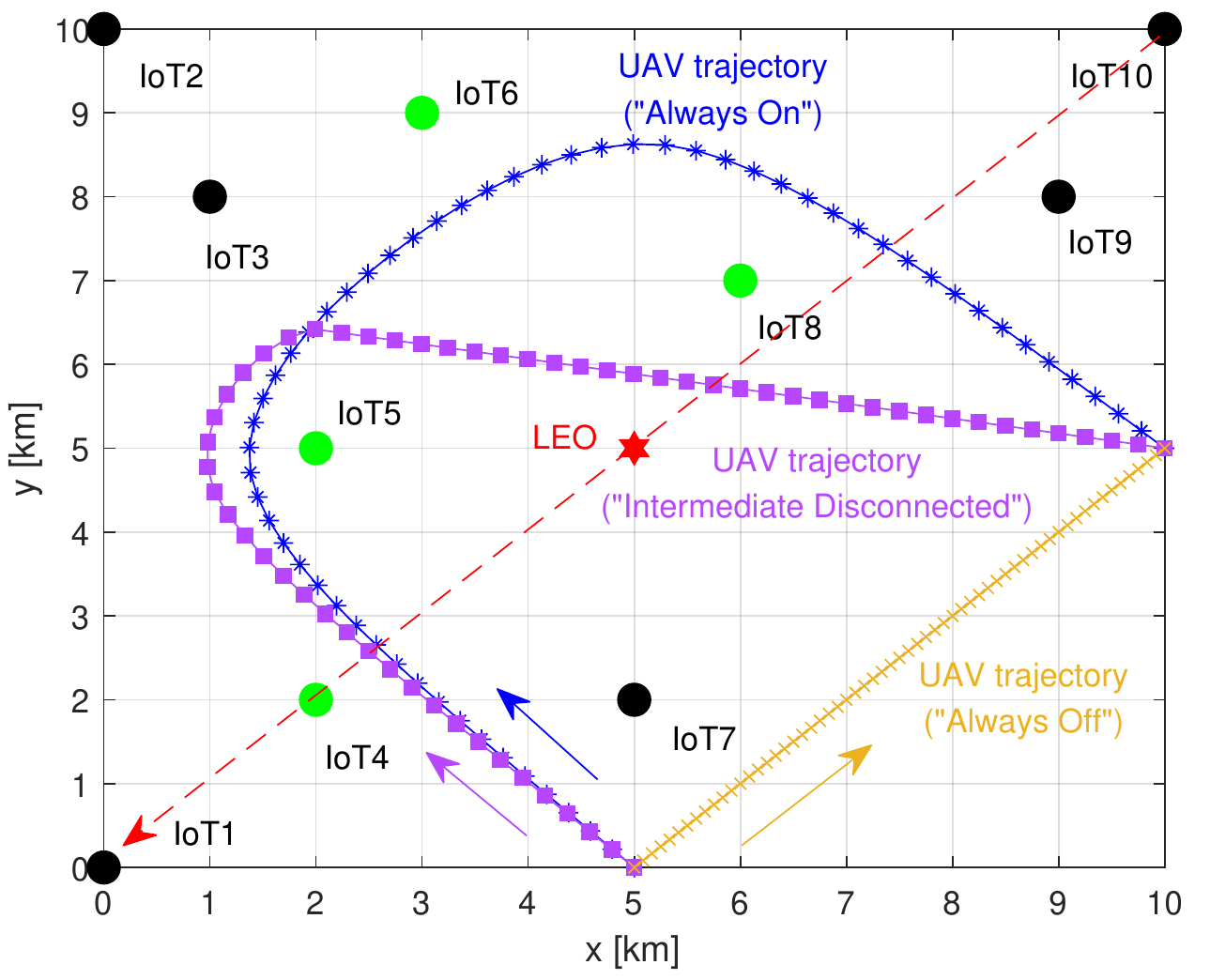}
                 \caption{Optimal UAV trajectories according to the different LEO access scenarios.}
         \label{fig:traj_new1}
     \end{figure}

\begin{figure}[t]
     \begin{subfigure}{\columnwidth}
         \centering
         \includegraphics[width=\columnwidth]{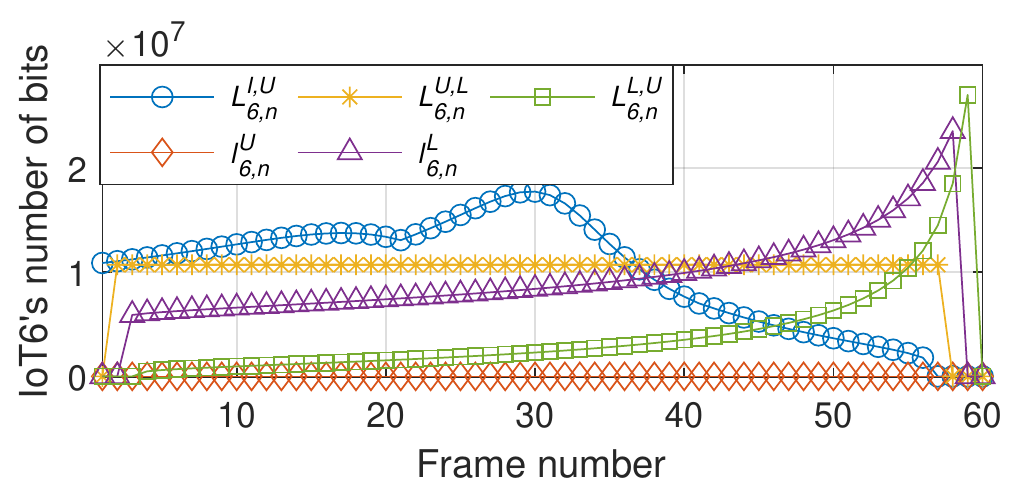}
         \caption{``Always On" scenario}
         \label{fig:bit_new1}
     \end{subfigure}
     \hfill
     \begin{subfigure}{\columnwidth}
         \centering
         \includegraphics[width=\columnwidth]{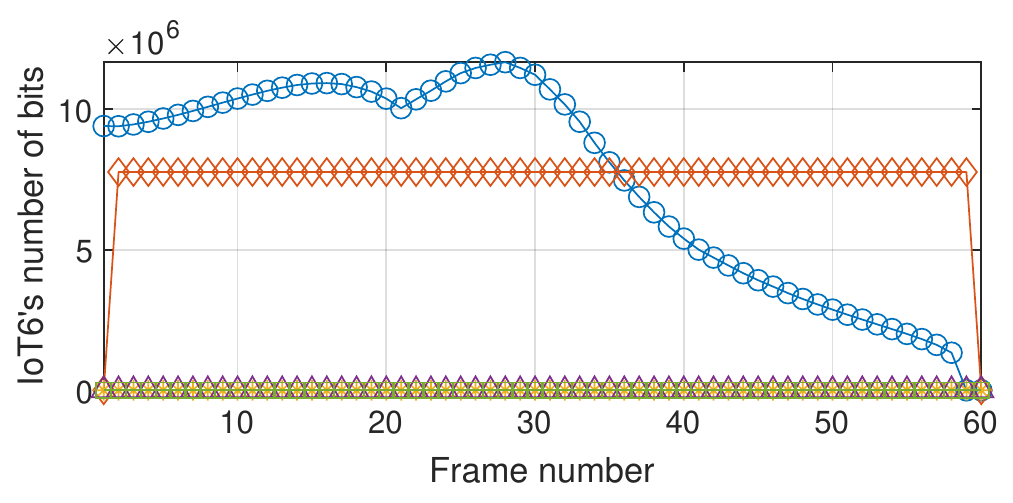}
         \caption{``Always Off" scenario}
         \label{fig:bit_new2}
     \end{subfigure}
          \hfill
     \begin{subfigure}{\columnwidth}
         \centering
         \includegraphics[width=\columnwidth]{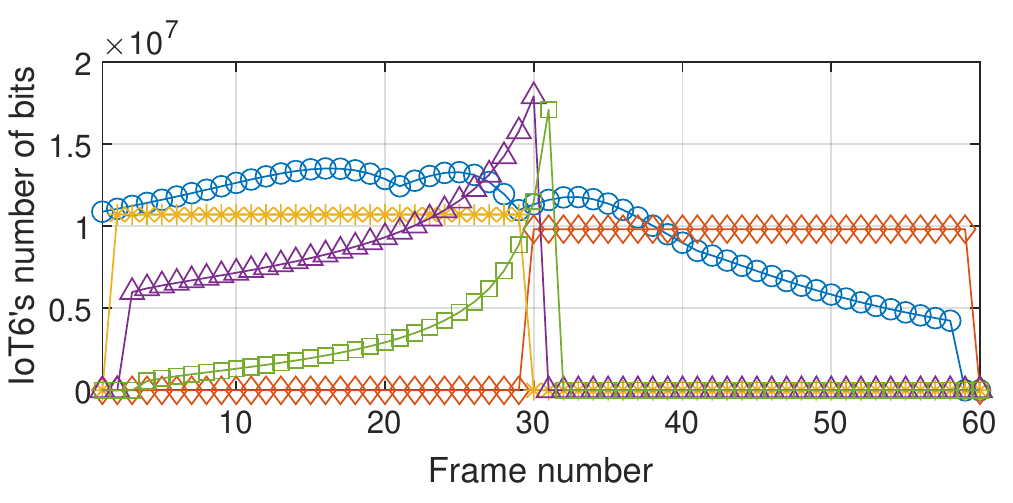}
         \caption{``Intermediate Disconnected" scenario}
         \label{fig:bit_new3}
     \end{subfigure}
        \caption{Optimal bit allocations for IoT sensor 6 in Fig. 4 according to the different LEO access scenarios.}
        \label{fig:traj_bit_new}        
\end{figure}

To better understand the proposed algorithms, Figs. 4 and 5 consider the partial optimization of UAV path planning or bit allocation. As shown in Fig. 4, there are $K = 10$ IoT sensors distributed randomly in a  $10 \ {\rm{km}}\, \times \,10 \ {\rm{km}}$ area within the beam coverage of the central LEO satellite, i.e., $\alpha_{k,n}=1$, for all $n \in \mathcal{N}$ and $k \in \mathcal{K}$. With the LEO visible time of $T_v = $ 830 s obtained from (\ref{equ:visible}) and a bandwidth of $B = $ 40 MHz [\ref{channel2}], the data size collected from each IoT sensor is randomly determined based on the computation capability of the UAV in (\ref{equ:w2}). In our simulation, the scheduling variable $\beta_{k,n}$ is defined from (\ref{equ:w2}), i.e., ${\beta _{k,n}} = [0 \ 0 \ 0\ 1\ 1\ 1\ 0\ 1\ 0\ 0]$, for $k \in \mathcal{K}$ and $n \in \mathcal{N}$, as shown in Fig. 4. The IoT sensors with ${\beta _{k,n}} = 0$ for UAV computing and with ${\beta _{k,n}} = 1$ for LEO computing are indicated by black-colored circles and green-colored circles, respectively, while the LEO satellite, indicated by a red-colored hexagram, travels along the red dotted line. The initial and final positions of the UAV are $\boldsymbol{p}_I^U= \left( {5,0,0} \right)$ to $\boldsymbol{p}_F^U = \left( {10,5,0} \right)$. 

Fig. 4 shows the optimized UAV trajectories with the fixed equal bit allocation according to the different LEO satellite access scenarios. For this experiment, the latency constraint is $T$ = 360 s with $N$ = 60 and $\Delta =$ 6 s. In the ``Always On" case, the optimized UAV trajectory, represented by a blue asterisk line, is designed to fly close to the IoT sensors with LEO computing until its final destination. This can significantly reduce the large amount of uplink communication energy consumption induced by the long distance between the LEO satellite and UAV. In the ``Always Off" case, where only UAV computing is considered, the UAV flies along a straight path to a destination, which is represented by a yellow crossed line. In this case, the flying energy consumption must be reduced to to minimize the total UAV energy due to the fixed computation bit allocation. In the ``Intermediate Disconnected" case, where the LEO communication is lost at $N_t = {{ {N} } \mathord{\left/ {\vphantom {{ {N} } 2}} \right. \kern-\nulldelimiterspace} 2}$, the optimized UAV trajectory, represented by a purple square line, tends to fly close to the IoT sensors with LEO computing for $n = 1, \cdot  \cdot  \cdot ,N_t$. Then, in the frame period of $n = N_t + 1, \cdot  \cdot  \cdot ,N$ where LEO communication is disconnected, the UAV flies straight to the final destination because it performs only UAV computing.

Fig. 5 illustrates the optimized bit allocations for IoT sensor 6 shown in Fig. 4 with the fixed constant-velocity UAV trajectory according to different LEO access scenarios. Except for the UAV trajectory, the simulation environment is the same as in Fig. 4. In Fig. 5(a), the optimal bit allocations $L_{k,n}^{I,U}$, $L_{k,n}^{U,L}$, $l_{k,n}^{L}$, $L_{k,n}^{L,U}$ by proposed Algorithm 1 are shown for LEO computing in the ``Always On" case. First, most of the uplink bits $L_{k,n}^{I,U}$ are allocated between frames 20 to 35, which corresponds to the period where the UAV flies closest to IoT sensor 6. The offloading bits $L_{k,n}^{U,L}$ are allocated equally in the entire frame because the equal bit allocation can achieve the minimal communication energy from (\ref{equ:UAV1}). Finally, the LEO computing bits $l_{k,n}^{L}$ and LEO downlink bits $L_{k,n}^{L,U}$ are mostly allocated in the latter parts between frames 50 to 60 to satisfy the inequality constraints of (\ref{st5}) and (\ref{st6}). In Fig. 5(b), the optimized bit allocations $L_{k,n}^{I,U}$ and $l_{k,n}^{U}$ obtained by proposed Algorithm 2 are shown for UAV computing of the``Always Off" case. Since the UAV cannot communicate with the LEO satellite, the computing process is entirely at the UAV-mounted cloudlet. The uplink bits $L_{k,n}^{I,U}$ and the computing bits $l_{k,n}^{U}$ are assigned the same as $L_{k,n}^{I,U}$ and $L_{k,n}^{U,L}$ in Fig. 5(a), respectively. However, $l_{k,n}^{U}$ is dramatically reduced to $8 \times 10^6$ per frame compared to $10 \times 10^6$, as illustrated in Fig. 5(a). This is because the amount of data exceeding the UAV computation capability is excluded from the UAV computing. Fig. 5(c) shows the optimization result of bit allocation attained by proposed Algorithm 3 in the ``Intermediate Disconnected" case. LEO computing is performed during the first half of frames, i.e., $n = 1, \cdot  \cdot  \cdot ,N_t$, while UAV computing is performed during the second half of frames, i.e., $n = N_t + 1, \cdot  \cdot  \cdot ,N$. The computing bits $l_{k,n}^{L}$ at LEO and the downlink bits $L_{k,n}^{L,U}$ are reduced in proportion to the reduced frame duration of LEO computing compared to those shown in Fig. 5(a). For UAV computing, there are more computing bits $l_{k,n}^{U}$ allocated at the UAV than those from the case in Fig. 5(b). This means that less data exceeds the computational capability of the UAV thanks to the LEO computing.

\begin{figure}[t]
        \includegraphics[width=\columnwidth]{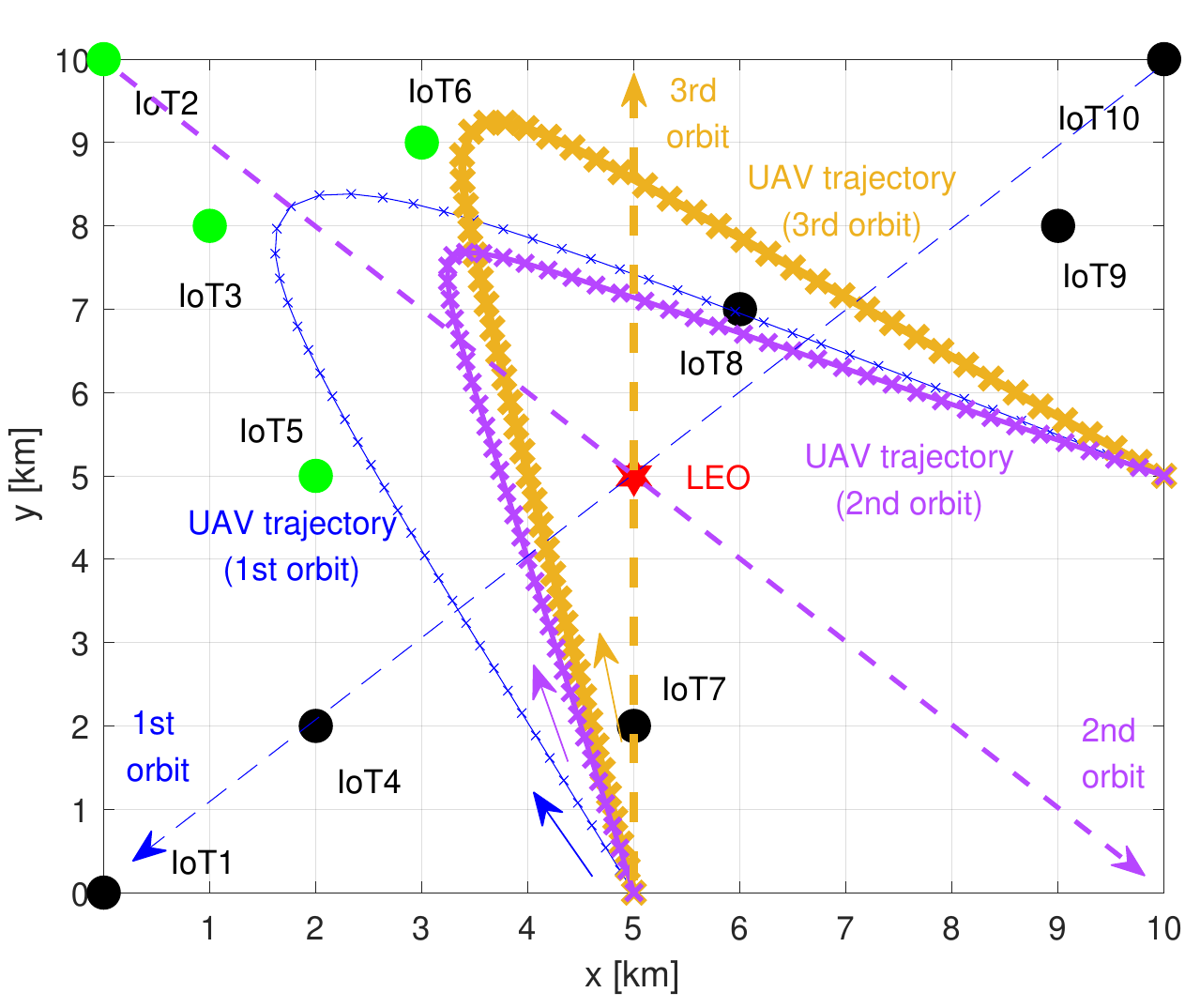}
        \caption{Optimal UAV trajectories according to different LEO satellite orbits, where the IoT sensors with LEO computing are deployed at the corner.}
        \label{fig:traj_new2}        
\end{figure}

\begin{figure}[t]
    \includegraphics[width=\columnwidth]{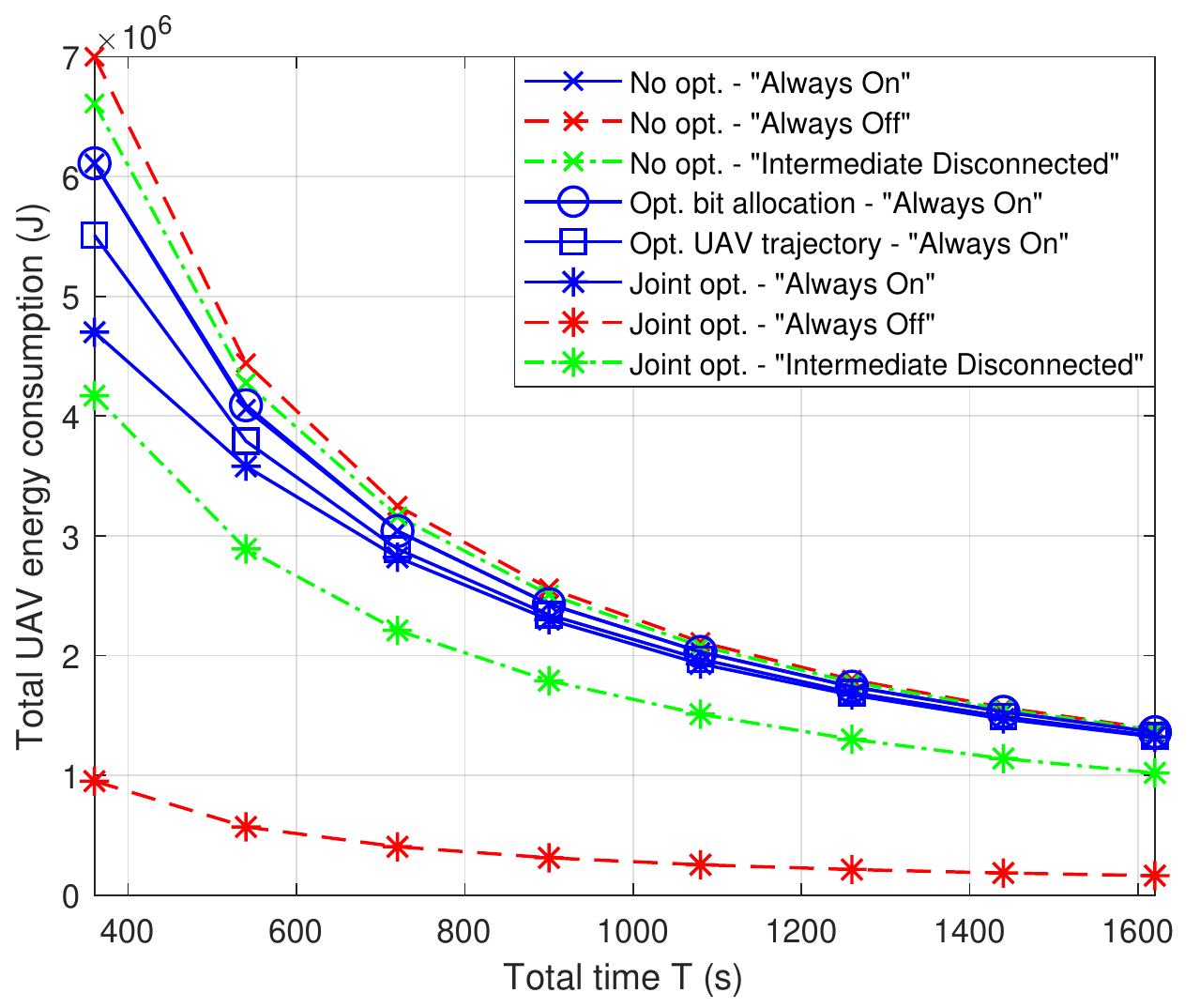}
    \caption{Comparison of the total UAV energy consumption for different optimization schemes in the three LEO satellite access scenarios.}
\end{figure}

Fig. 6 shows the optimal UAV trajectories according to the different LEO satellite orbits in the ``Always On" scenario, where the IoT sensors that need LEO computing are clustered at the corner, i.e., ${\beta _{k,n}} = [0\ 1\ 1\ 0\ 1\ 1\ 0\ 0\ 0\ 0]$, for $k \in \mathcal{K}$ and $n \in \mathcal{N}$. In this deployment, the three different movements of the LEO satellite in different orbital directions are considered. In the first orbit moving from the upper right corner to the lower left corner, the UAV flies near the corner area with IoT sensors with LEO computing to its final destination. In the second orbit moving from the upper left corner to the lower right corner, the UAV flies in a diagonally downward direction along its own orbit rather than the optimized UAV trajectory for the first orbit. In the third orbit moving upwards from below the midpoint, the UAV flies in an upward direction along its own orbit rather than the optimal UAV trajectory  for the first orbit. From these results, we can see that the LEO movements resulting from the orbit influences the optimal UAV path so as to reduce the communication energy consumption between the UAV and the LEO satellite.

Fig. 7 compares the total UAV energy consumption of the joint optimization scheme with reference schemes in three LEO satellite access scenarios. For this experiment, the latency constraint is $T$ = [360:90:1620] s with $N$ = [60:15:270] and $\Delta =$ 6 s, while the remaining simulation parameters are the same as in Figs. 4 and 5. First, the no optimization scheme consumes the highest energy in the three scenarios, among which the largest energy consumption takes place in the ``Always Off" case, where only the UAV computing is performed. This is natural since the UAV-mounted cloudlet has a slightly larger burden in terms of the energy consumption with no support of the LEO. In the ``Always On" case, for $T=360$ s, the total UAV energy consumption for the joint optimization scheme is the lowest at $4.6 \times 10^6$ J, whereas the optimized UAV trajectory scheme with fixed equal bit allocation requires $5.5 \times 10^6$ J, and the optimized bit allocation with the constant-velocity UAV and no optimization schemes requires $6.1 \times 10^6$ J. This implies that the UAV path planning is more effective in terms of UAV energy consumption than bit allocation. Moreover, the total energy consumption in all schemes decreases as the total time increases. This is because the same amount of data is processed over a longer period of time. Compared to the total UAV energy consumption of the joint optimization scheme in the ``Always Off" scenario, those of the joint optimization scheme in other scenarios are much higher since the UAV flies straight to its final destination when the LEO satellite connection is lost, as in Fig. 4. However, there is a trade-off between the total UAV energy consumption and the collected data usage rate for computing, which determines the amount of data executed at cloudlet, which is analyzed in the following figure.

\begin{figure}[t]
    \includegraphics[width=\columnwidth]{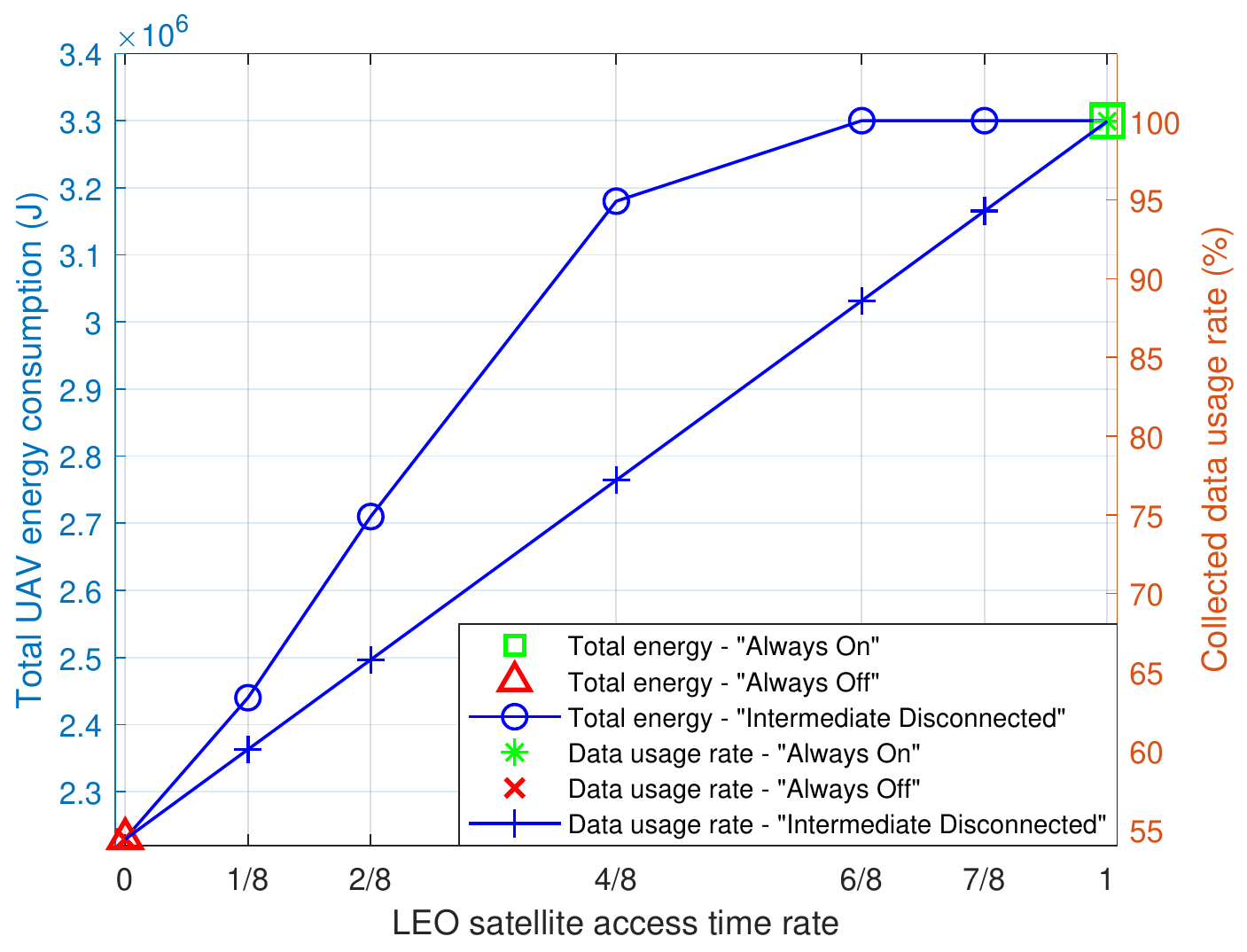}
    \caption{Relationship between the total UAV energy consumption and the collected data usage rate in three LEO satellite access scenarios according to the LEO satellite access time rate.}
\end{figure}

Fig. 8 shows the relationship between the total UAV energy consumption and the collected data usage rate for computing in the different LEO accessibility scenarios. Any amount of data exceeding the UAV computation capability is excluded from UAV computing. For this experiment, the scheduling variables are defined as ${\beta _{k,n}} = [0 \ 0 \ 1\ 1\ 1\ 1\ 0\ 1\ 0\ 0]$, for $k \in \mathcal{K}$ and $n \in \mathcal{N}$. The UAV computation capability is applied to 226 Mbits by using the CPU frequency at the UAV server ${f_n^U}$ = $9.75 \times {10^9}$ cycles/s. In the ``Always On" scenario, the LEO satellite access time rate is 1. At this time, the total UAV energy consumption is $3.3 \times 10^6$ J and the collected data usage rate is $100 \% $. In the ``Always Off" case, where the LEO satellite access time rate is 0, the total UAV energy consumption is $2.24 \times 10^6$ J and the collected data usage rate is $54 \% $. Although the energy consumption in the ``Always Off" case is dramatically reduced, the utilization rate of the collected data is also cut in half. In the ``Intermediate Disconnected" case, as the LEO satellite access time rate increases, the total UAV energy consumption and the collected data usage rate increase differently. When the LEO satellite access time rate is above $6/8$, the total UAV energy consumption is saturated with the total UAV energy consumption of the ``Always On" case. This is because the straight flight segment of the UAV to the final destination after disconnecting with the LEO satellite matches that of the ``Always On" case. Also, when the LEO satellite access time rate is more than 7/8, the collected data usage rate is more than about $95 \%$. In this simulation environment, adequate data usage and energy consumption is achieved with more than a 7/8 LEO satellite access time rate.

\section{Conclusions}

In this paper, a marine IoT system using hybrid LEO and UAV computing for real-time utilization of marine data has been analyzed according to the different LEO satellite access scenarios: ``Always On,” ``Always Off” and ``Intermediate Disconnected”. For each scenario, we proposed the joint optimization problem of bit allocation for computing and communication in offloading and UAV path planning to minimize the total UAV energy consumption under latency, energy budget, and UAV operational constraints. To solve the optimization problem, we developed an SCA-based algorithm whose performance in terms of energy efficiency was validated via numerical results compared to conventional approaches with partial optimization that design only the bit allocation or UAV trajectory. According to LEO satellite access time and its orbit direction, the path planning of the UAV is optimized differently for energy saving, whose impact is pronounced for the case when the LEO connectivity is unstable or disconnected. In future works, different existing LEO deployments should be further considered with various heights of multiple satellites and UAVs.

\ifCLASSOPTIONcaptionsoff
  \newpage
\fi

\end{document}